%% file: EXO-11-013_temp.tex
\pdfoutput=1

\documentclass[11pt,twoside,a4paper,cmspaper,final,collab]{cms-tdr}
\def\svnVersion{62105:62106}\def\cmsCernNo{2011-064}\def\cmsCernDate{\today}\def\cmsMessage{Submitted to the Journal of High Energy Physics}

\begin{document}\cmsNoteHeader{EXO-11-013}

\hyphenation{had-ron-i-za-tion}
\hyphenation{cal-or-i-me-ter}
\hyphenation{de-vices}
\RCS$Revision: 62105 $
\RCS$HeadURL: svn+ssh://alverson@svn.cern.ch/reps/tdr2/papers/EXO-11-013/tags/jhep-1/EXO-11-013.tex $
\RCS$Id: EXO-11-013.tex 62105 2011-06-17 20:31:30Z alverson $
\cmsNoteHeader{EXO-11-013} 
\title{Search for Light Resonances Decaying into Pairs of Muons as a Signal of New Physics}

\address[tamu]{Texas A\&M University}
\author[tamu]{J. Pivarski, A. Safonov, A. Tatarinov}
\author[cern]{The CMS Collaboration}

\date{\today}

\abstract{
A search for groups of collimated muons is performed using a data sample collected  by the CMS experiment at the LHC, at a centre-of-mass energy of 7 TeV, and corresponding to an integrated luminosity of 35 pb$^{-1}$.  The analysis searches for production of new low-mass states decaying into pairs of muons and is designed to achieve high sensitivity to a broad range of models predicting leptonic jet signatures. With no excess observed over the background expectation, upper limits on the production cross section times branching fraction times acceptance are set, ranging from 0.1 to 0.5~pb at the 95\% CL depending on event topology. In addition, the results are interpreted in several benchmark models in the context of supersymmetry with a new light dark sector exploring previously inaccessible parameter space.}

\hypersetup{%
pdfauthor={CMS Collaboration},%
pdftitle={Search for Light Resonances Decaying into Pairs of Muons as a Signal of New Physics},%
pdfsubject={CMS},%
pdfkeywords={CMS, physics}}

\maketitle 

\input{intro.tex}
\input{selections.tex}
\input{mass_analysis.tex}
\input{conclusions.tex}

\bibliography{auto_generated}   
\cleardoublepage\appendix\section{The CMS Collaboration \label{app:collab}}\begin{sloppypar}\hyphenpenalty=5000\widowpenalty=500\clubpenalty=5000\input{EXO-11-013-authorlist.tex}\end{sloppypar}
\end{document}

%% file: intro.tex
\section{Introduction}
\label{sec:intro}

Recent astrophysical observations of an excess of high-energy positrons in the cosmic-ray spectrum~\cite{Pamela-positron} have motivated the rise of new physics scenarios~\cite{Arkani-Hamed} suggesting that this excess may be associated with annihilations of dark matter particles  in the galactic halo. These models may also accommodate the observed discrepancies in direct searches for dark matter~\cite{Dama,Finkbeiner,Alves}. One realization of such models assumes an extra $U(1)$ gauge symmetry with weak coupling to the standard model (SM). The $U(1)$ symmetry is broken, leading to a light vector boson (with mass $m \sim O(1$~GeV/$c^2$)), referred to as a ``hidden'' or ``dark'' photon $\gamma_{\rm dark}$, which can have a small kinetic mixing with the SM photon providing a portal for the hidden-sector photon to decay into leptons and, if kinematically allowed, hadrons. More complex models can lead to a whole hierarchy of hidden-sector states or can have dark photons preferentially couple to leptons~\cite{mediator-lepton-coupling2}. Hidden sectors can be realized naturally in supersymmetric (SUSY) models  where coupling of the dark sector to the SUSY sector can be enhanced. At the Large Hadron Collider (LHC), if SUSY exists and is kinematically accessible, these models predict production of dark photons as part of the SUSY cascades. The new light hidden states may be produced in decays of the lightest supersymmetric particle (sparticle) of the visible sector. In the extensions of the Minimal Supersymmetric Standard Model (MSSM) with the lightest neutralino being the lightest SUSY particle (LSP) in the visible sector, the MSSM LSP can decay to the light hidden-sector particles and sparticles, while a heavy dark fermion provides a cold dark matter candidate. Alternatively, the lightest MSSM neutralino can decay to light hidden particles and a heavier dark neutralino~\cite{BaiHan}, which becomes a cold dark matter candidate. Because the MSSM LSP in these models is unstable and is not subject to the constraints on a cold dark matter candidate, there are scenarios~\cite{Ruderman} where the LSP is a squark decaying into a quark and the light hidden-sector states. Depending on the complexity of the light dark sector, at the LHC one may expect either a single dark photon at the end of each SUSY cascade or a whole cascade of hidden-sector-state decays with emission of multiple dark photons. Subsequent decays of new states into leptons lead to appearance of energetic collimated groups of leptons, the characteristic ``leptonic jet'' signature~\cite{Strassler}. While the spectrum of such leptonic jets depends on masses of sparticles that are part of the SUSY cascade in which they are produced, typical scenarios accessible at the LHC predict leptonic jets with transverse momenta ranging from tens to several hundreds of GeV/$c$.

Previous searches for low-mass dilepton resonances have been performed at the Tevatron~\cite{D01,D02}, as well as Belle~\cite{Belle}, BaBar~\cite{babar-low-ma}, and LEP~\cite{Aleph}, and revealed no signals of new physics. Due to the large cross section of SUSY production via strong interaction, the LHC may provide access to these new states with early data motivating the search for anomalous production of collimated groups of leptons. 

This paper describes a topology-based search for new light resonances decaying to pairs of muons using data collected by the Compact Muon Solenoid (CMS) experiment during the 2010 LHC data-taking period and corresponding to an integrated luminosity of $35.0\pm1.4$ pb$^{-1}$. This analysis searches for events with one or more muon pairs that are consistent with being produced in the decays of the same particle type. Assuming on-shell production of at least a fraction of these new bosons per event, the new physics would manifest itself as an enhancement in the production rate of muon pairs consistent with a certain common mass. The results of the search are presented in a model-independent fashion, limits are also set on specific benchmark scenarios~\cite{Ruderman,BaiHan} in the context of SUSY models.

%% file: selections.tex
\section{Detector, Dataset, and Trigger}

The CMS detector is a general-purpose apparatus providing excellent momentum and direction measurement of particles produced in \Pp\Pp\ collisions at the LHC. The central feature of CMS is a superconducting solenoid of 6~m internal diameter. Within the volume of a 3.8 T magnetic field are the silicon pixel and strip tracker, the crystal electromagnetic calorimeter 
and the brass/scintillator hadron calorimeter. 
Muons are measured in gas-ionization detectors embedded in the steel return yoke. In addition to the barrel and endcap detectors, CMS has extensive forward calorimetry. CMS uses a right-handed coordinate system, with the origin at the nominal interaction point, the $x$ axis pointing to the centre of the LHC ring, the $y$ axis pointing up (perpendicular to the LHC plane), and the $z$ axis along the counterclockwise beam direction. The polar angle $\theta$ is measured from the positive $z$ axis and the azimuthal angle $\phi$ is measured in the $x$-$y$ plane. The pseudorapidity 
$\eta = -\ln [\tan (\theta/2) ]$  is frequently used instead of the polar angle $\theta$. Here we only briefly describe the components of the CMS directly relevant to this analysis; the full details of the detector, its subsystems, and performance are described elsewhere~\cite{CMS}. 

The inner tracker measures charged particles within the pseudorapidity range $|\eta| < 2.5$. It consists of 1440 silicon pixel and 15\,148 silicon strip detector modules and is located inside the superconducting solenoid. It provides an impact parameter resolution of $\sim$\,15~$\mu$m and a transverse momentum ($p_{T}$) resolution of about 1.5\,\% for 100~GeV/$c$ particles. The muons are measured in the pseudorapidity window $|\eta|< 2.4$, with detection planes made of three technologies: Drift Tubes, Cathode Strip Chambers, and Resistive Plate Chambers. Matching the muons to the tracks measured in the silicon tracker results in a transverse momentum resolution between 1 and 5\,\%, for $p_{T}$ values up to 1~TeV/$c$. 

Because of the high rate of the collisions, the CMS uses a two-level dedicated trigger system. The first level (L1) of the CMS trigger, composed of custom hardware processors, is designed to select, in less than 1~$\mu$s, the most interesting events, using information from the calorimeters and muon detectors. The High Level Trigger (HLT) processor farm, running a simplified and highly optimized version of the CMS offline reconstruction, is designed to decrease further the event rate 
down to a maximum of 300~Hz, before data storage.

The data used in this analysis have been collected with the inclusive muon triggers with the lowest available $p_T$ threshold. At level 1, the data are selected using muon candidates reconstructed by the L1 muon hardware, followed by the confirmation at the HLT, where muons are reconstructed by matching tracks reconstructed in the muon system with the tracks reconstructed in the silicon tracker detectors to refine the muon $p_T$ measurement. Because the trigger configuration was changing during the data taking period, there are three distinct parts of the dataset where the triggers used had transverse momentum thresholds of 9, 11, and 15~GeV/$c$ at the HLT. In all cases, level 1 muon thresholds were low enough to ensure that the HLT thresholds are at the plateau of the L1 muon efficiency. To make the selected data uniform, we additionally require offline events to contain at least one trigger candidate with $p_T>15$~GeV/$c$ as measured online, such that the final dataset is the same as it would have been if collected using a single inclusive muon trigger with $p_T>15$~GeV/$c$.

%% file: mass_analysis.tex
\section{Offline Selections and Analysis of the Data}

In this analysis we search for evidence of new light bosons decaying into pairs of muons. The new particles can be isolated or produced in groups, coming from cascade decays in the hidden sector ending with several instances of the lowest-state particle. In addition to dimuon decays, the new particle decay channels can include pairs of electrons and perhaps hadrons. In many scenarios, multiple instances of such a new boson can be produced per event. This analysis therefore searches for events with one or more muon pairs that are consistent with being produced in the decays of the same particle type. Assuming on-shell production of at least a fraction of these new bosons per event, the new physics would manifest itself as an enhancement in the production rate of muon pairs consistent with a certain common mass.

This analysis aims to reach two goals. First, to achieve high sensitivity for a representative range of specific new physics scenarios leading to characteristic muon-jet signatures and, second, to present results that would allow future interpretation in the context of other models of new physics yielding leptonic jet signatures. Essentially all classes of models of new physics relevant to this analysis lead to the production of events of several different topologies in terms of the number of collimated muon jets and multiplicities of muons within each jet. Because these topologies have different sources and levels of SM backgrounds, we categorize events using topologies with different expected signal-to-background ratios to maximize the overall sensitivity of the analysis.

\subsection{Data Selections and Reconstruction}
Offline, events are required to have at least one primary vertex
reconstructed in the  luminous region along the beamline to minimize background events not originating
from collisions. Selected events are further required to have at least one high-quality muon candidate, with $p_T>15$~GeV/$c$
matching the muon selected by the online trigger and within $|\eta|<0.9$ reconstructed using
an ``inside-out'' algorithm, thus ensuring high efficiency in the environment with multiple nearby muons.
This algorithm extrapolates silicon tracker tracks into the muon system and attaches to them individual tracklets (stubs)
reconstructed in muon chambers. Any stub in the muon
system can only be associated with one extrapolated tracker track most compatible with the stub. The
reconstructed muon candidate is required to have stubs in at least two out of four muon stations it crosses. To
be classified as high-quality, muon candidates are required to have at least eight hits in the silicon tracker. The
requirement of $|\eta|<0.9$ ensures high and well understood trigger
efficiency insensitive to the presence of muon hits from other nearby muons
expected in the signatures with collimated muons. It avoids the endcap region where the trigger
efficiency can be substantially diminished in the presence of multiple closely spaced muons because of the features of
the trigger electronics setup. Additional muon candidates are required to have $p_T>5$~GeV/$c$, to be contained
within $|\eta|<2.4$, and to satisfy the same quality requirements.

Muon jets are reconstructed by iteratively clustering muon candidates starting with the highest $p_T$. Each additional muon is added to the jet if the invariant mass of this muon and any
oppositely charged muon already assigned to the jet satisfies $m_{\mu\mu}<9$~GeV/$c^2$ and is
compatible with originating from the same vertex (confidence level of the vertex fit $>$ 1\%). The clustering procedure always converges and is independent of the order in which muons are added to muon jets. The choice of $m_{\mu\mu}<9$~GeV/$c^2$ ensures that muons originating from the same \cPqb\ quark are always clustered into the same muon jet and most muons originating from different \cPqb\ quarks are clustered into different muon jets.  It is also appropriate
for topologies predicted by most relevant models of new physics, as the typical masses of the heavier
hidden states originating the cascades are of the order of a few GeV/$c^2$. Note that the efficiency of this
clustering algorithm does not depend on the boost of the muon jet, thereby reducing the sensitivity of the analysis
to those details of the kinematics that can differ from one model to another. As a consequence of the clustering algorithm, each muon jet must contain at least one muon of each charge, but can contain arbitrarily many muons.
Within each muon jet, we identify ``fundamental dimuons,'' the pairs of oppositely charged muons that are most likely to have arisen from individual dark photon decays. Since all dark photons in the event have the same mass, the assignment is performed by selecting a combination of pairs that yields minimal difference in dimuon mass among the pairs.

\subsection{Event Categorization}
All events are categorized according to the number of muon jets $N$ and the number of muon candidates $n_i$ in the $i^{\mbox{\scriptsize th}}$ muon jet, thus forming topologies denoted as $R^{N}_{n_1 ... n_N}$. No isolation requirements are imposed on any of the muon candidates. Muons that do not belong to any muon jet (which may arise from SUSY cascades, rather than dark photon decays) are neither used to identify nor to reject signal events. The event selection is also insensitive to the presence of hadronic jets or missing transverse energy in the event.

In topologies with multiple dimuon candidates originating from the same particle type, the reconstructed masses of all fundamental dimuons would be consistent with each other within detector resolution for signal events, but not necessarily for backgrounds. Therefore, we build a $K$-dimensional distribution of reconstructed dimuon masses $m_1, ..., m_K$ for each topology, where $K$ is the number of reconstructed dimuons per event. The signal of new physics would appear as an enhancement of events at a point near the $K$-dimensional diagonal with $m_1 \approx ...\approx m_K \approx m_0$, where $m_0$ is the mass of the new particle. While the distributions of background events are not smooth because of the low-mass resonances, they extend beyond the diagonal in a known way. The only exception is the $R^1_2$ topology with exactly one fundamental dimuon per event: the signal would appear as a narrow peak in the 1D distribution of dimuon mass $m$. For topologies with $K$ dimuon candidates, we define the signal region as a ``corridor'' near the diagonal in the $K$-dimensional space of width $5 \times \sigma(m_{\mu\mu})$, where $\sigma(m_{\mu\mu}) = 0.026\;{\rm{GeV}}/c^2 + 0.0065\, m_{\mu\mu}$ for barrel ($|\eta^{\mu\mu}| < 0.9$, where $\eta^{\mu\mu}$ is the psedorapidity of the dimuon momentum defined as a vector sum of momenta of the muons in the dimuon) and $0.026\;{\rm{GeV}}/c^2  + 0.013\, m_{\mu\mu}$ for endcap ($0.9 < |\eta^{\mu\mu}| < 2.4$). The parameterization for $\sigma(m_{\mu\mu})$ was derived from studies of $\JPsi$, $\psi'$, $\phi$, and $\rho/\omega$
resonances, as well as high-$p_T$ Monte Carlo simulations, and corresponds to the resolution expected of hypothetical dimuons with $p_T \approx 300$~GeV/$c$ in the barrel region and $p_T \approx 150$~GeV/$c$ in the endcap. Better mass resolution for lower momentum dimuons makes this definition of the signal corridor conservative for $p_T^{\mu\mu}<300 \;(150)$ GeV/$c$ in the barrel (endcap). Dimuon momentum spectum predicted in typical models rarely extends beyond 300-400 GeV/$c$ making this choice acceptable for the entire range of expected dimuon momenta. After the shape of the background events distribution in the $K$-dimensional space is measured, the data in the off-diagonal part can be used to obtain the background normalization, which can then be used to fit the data in the near-diagonal region for signal plus background.

If an enhancement were to be observed in the diagonal regions, one could further construct the invariant mass of combinations of dimuons in the same muon jet to search for possible structure, e.g. a process of the type $a_2 \to a_1 a_1 \to (\mu\mu) (\mu\mu)$ would lead to an enhancement in the invariant mass of pairs of dimuons consistent with $m(a_2)$. To maintain analysis sensitivity for models with $m(a_2)<2m(a_1)$ where one of both $a_1$ bosons are produced off-shell, events with high multiplicity muon jets, e.g. topology $R^1_4$, are to be analyzed for evidence of an enhancement in the ``quadmuon'' invariant mass distribution.

\begin{figure}[tbh]
\centering
\begin{tabular}{cc}
\includegraphics[width=0.45\linewidth]{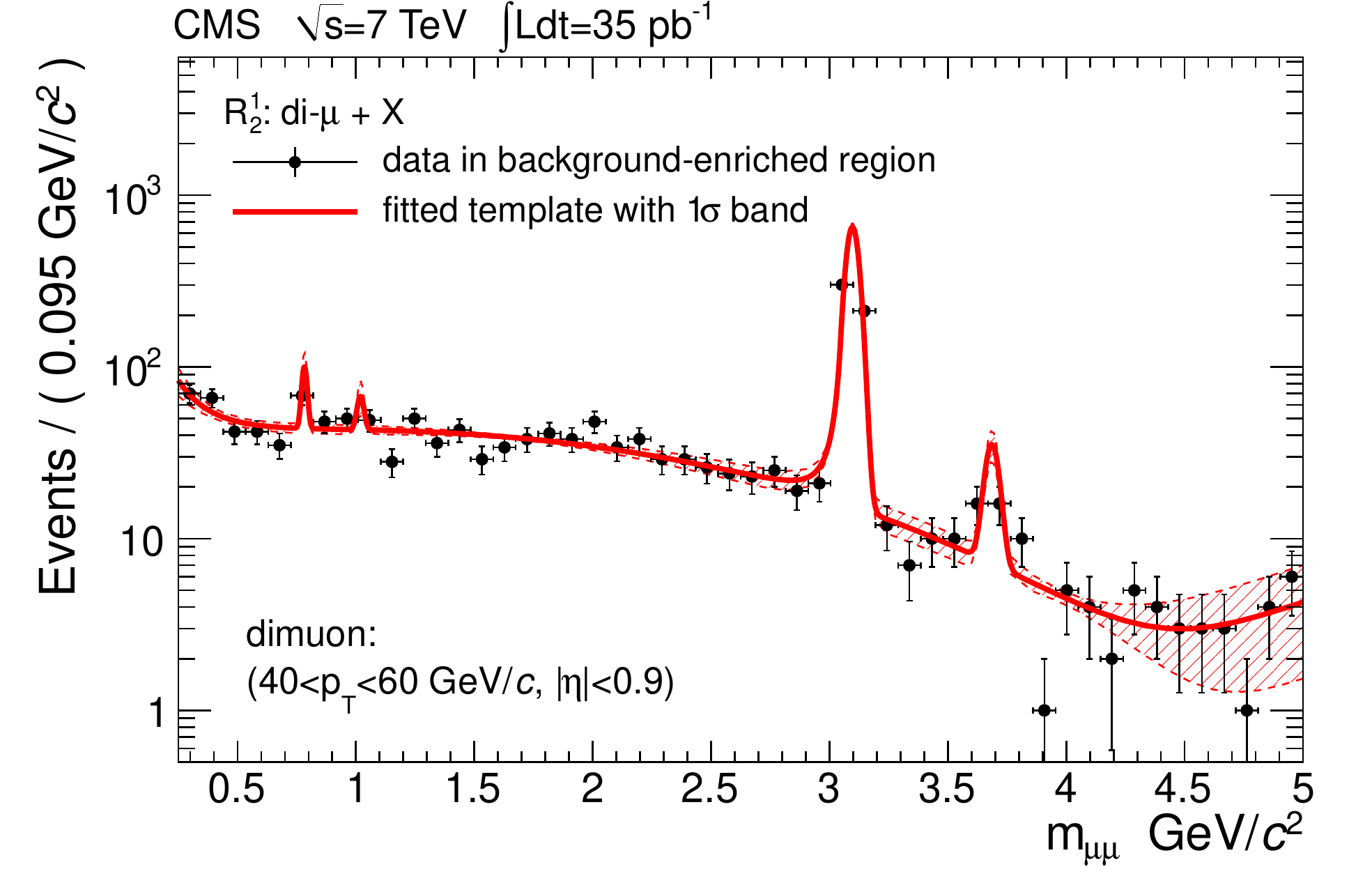}     \put(-40,90){\large (a)} &
\includegraphics[width=0.45\linewidth]{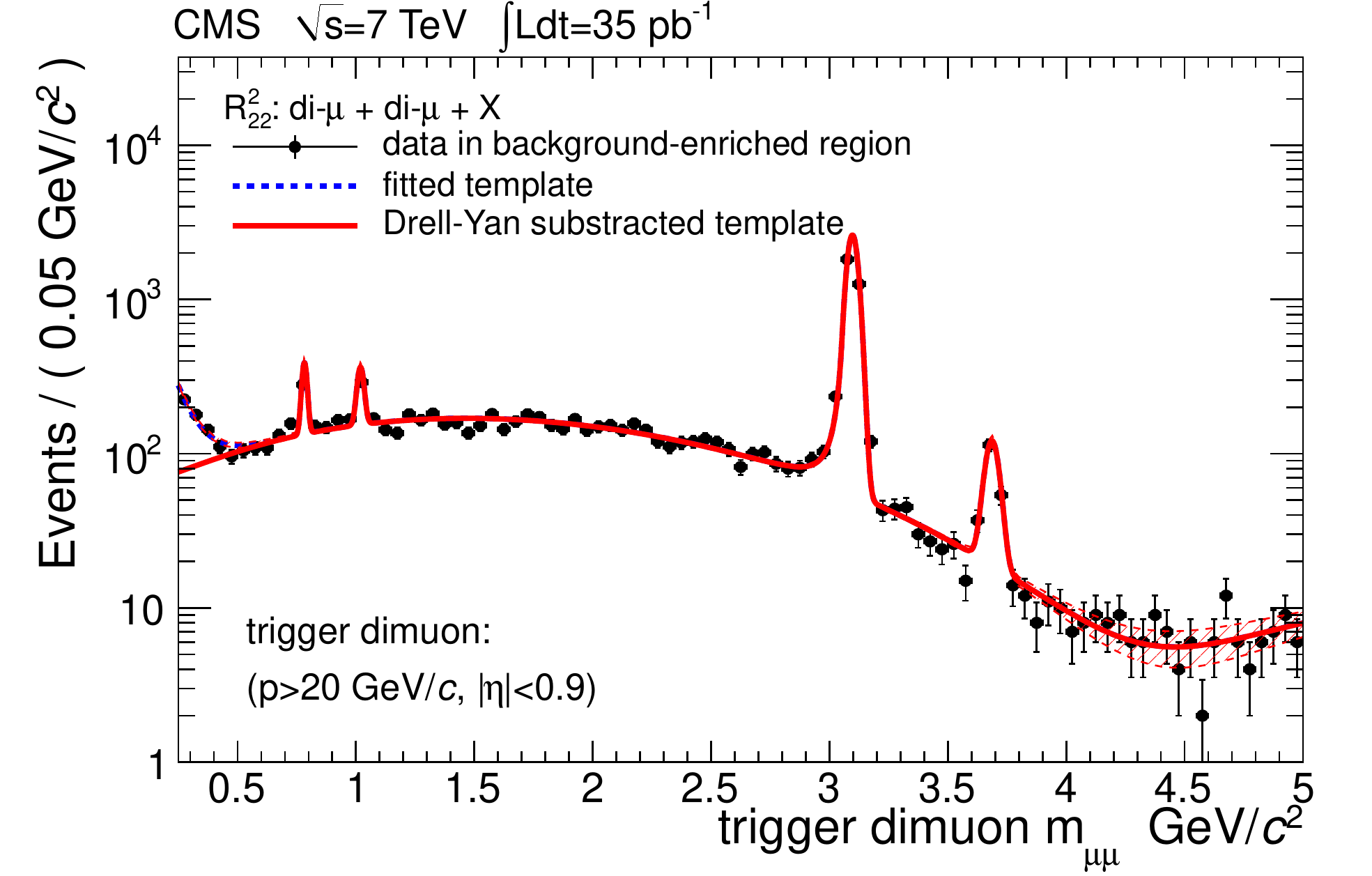}      \put(-40,90){\large (b)}\\
\includegraphics[width=0.45\linewidth]{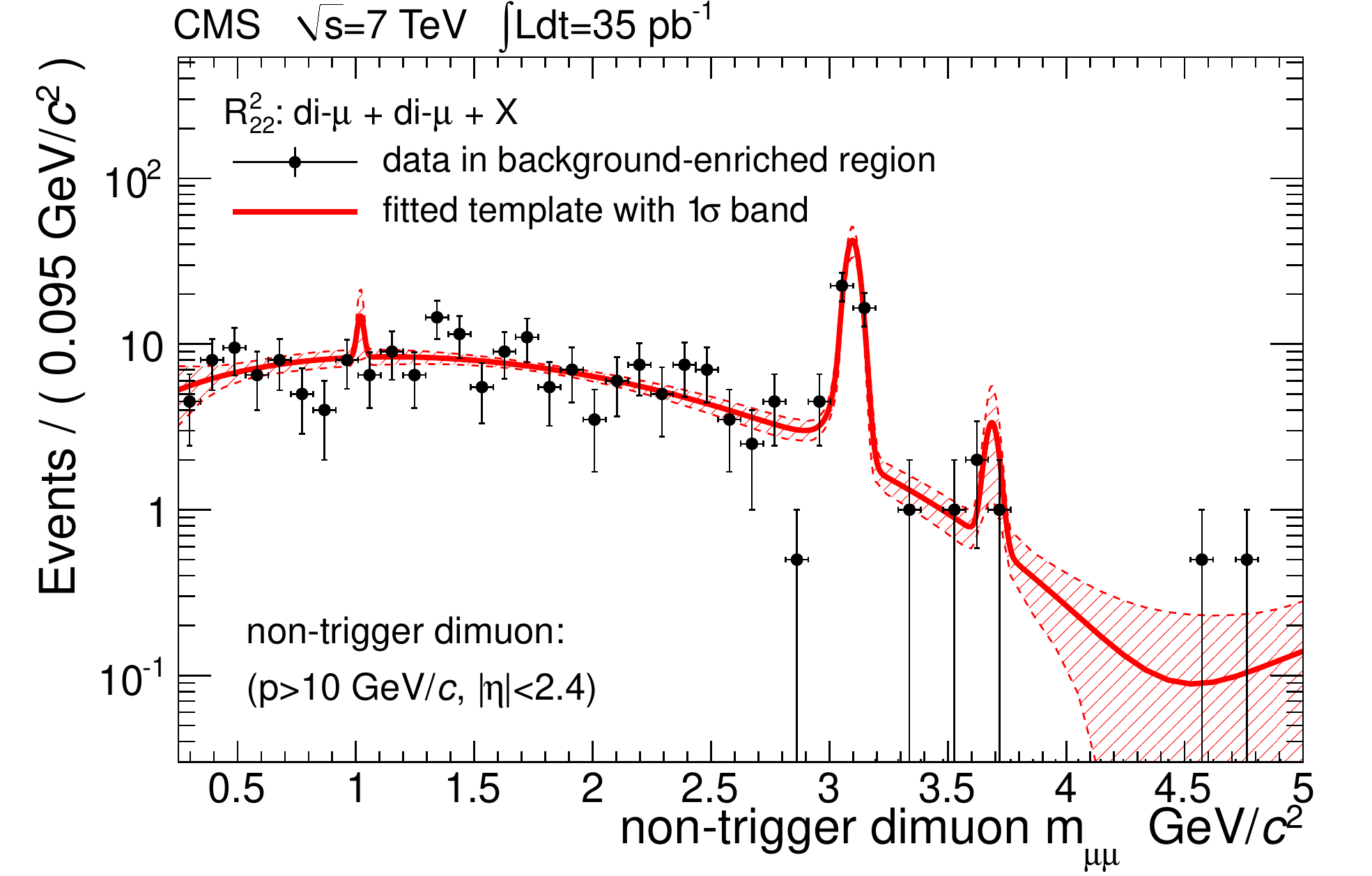}     \put(-40,90){\large (c)}&
\includegraphics[width=0.45\linewidth]{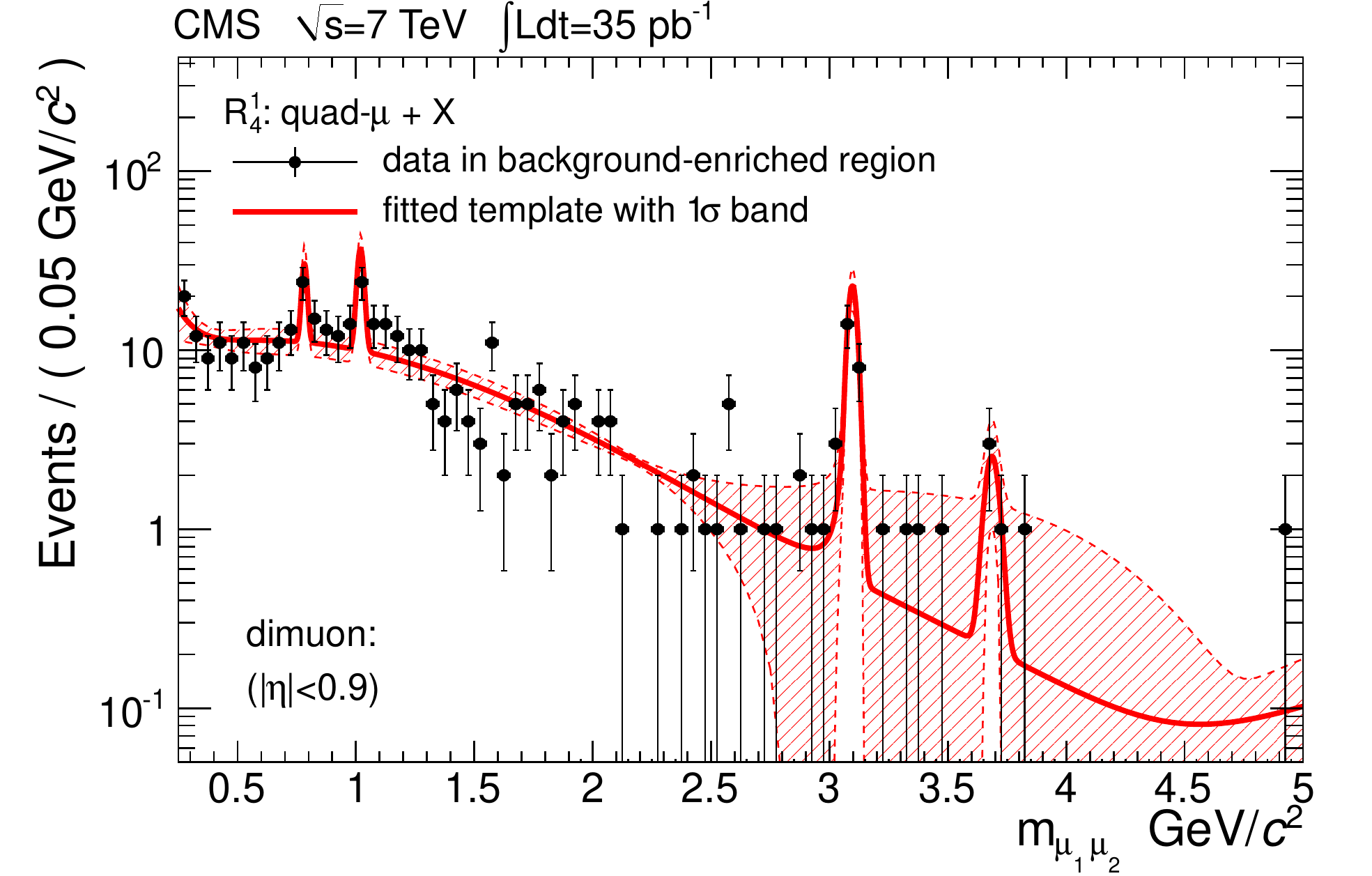}     \put(-40,90){\large (d)}\\
\includegraphics[width=0.45\linewidth]{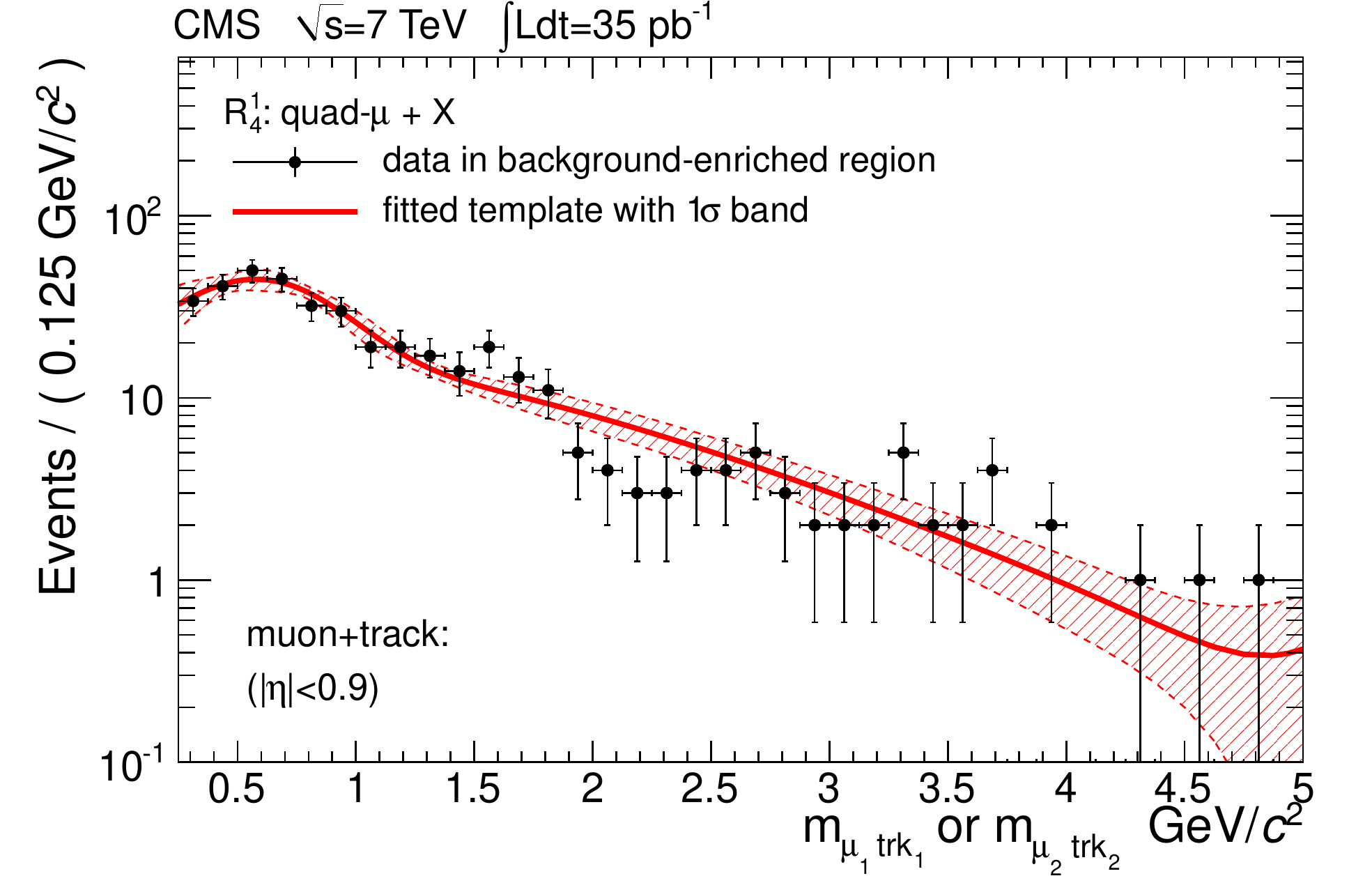}      \put(-40,90){\large (e)}&
\includegraphics[width=0.45\linewidth]{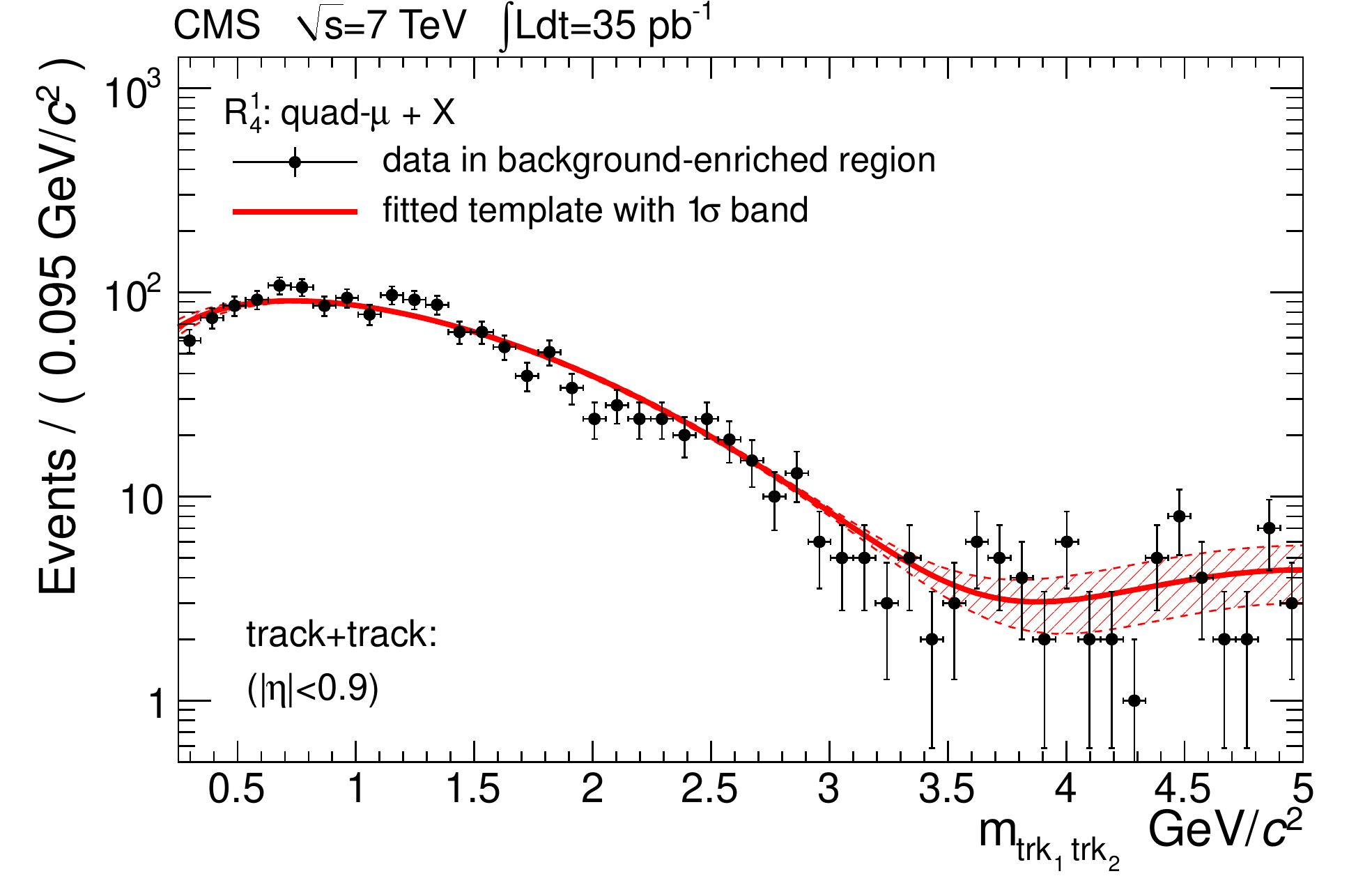}      \put(-40,90){\large (f)}\\
\end{tabular}
\caption{Dimuon invariant mass distributions in the background-enriched samples for different topologies. The data are overlaid with parameterized functions, fitted to the data, and used to construct mass-shape templates for the distributions of background events in signal regions.
\label{fig:background_distributions}}
\end{figure}

\subsection{Background Estimation}
The SM background contributions to event topologies $R^{N}_{n_1 ... n_N}$  vary significantly in terms of the overall rates and the composition of contributing processes. The single-dimuon topology $R^1_2$ suffers from a particularly high rate of the SM backgrounds because of \cPqb\cPaqb\ and Drell-Yan processes. Without additional selections, the SM backgrounds would be too large to maintain sensitivity to signals with picobarn-scale cross sections. To reduce the SM backgrounds, events in the $R^1_2$ topology are additionally required to have the muon jet transverse momentum satisfy $p_T^{\mu\mu}>80$~GeV/$c$. This requirement dramatically improves the sensitivity of this topology to new physics signals that predict highly boosted muon jets. At the same time this requirement reduces acceptance for signal events containing only one dimuon per event, particularly for models, in which the muon jets have lower boosts. For such models the sensitivity is driven by events with topologies containing two or more dimuons per event, for which no high momentum requirement is imposed. Data reconstructed in topology $R^1_2$ with lower momentum dimuon candidates are used for background studies and validation of the background estimation techniques. The other special case is topology $R^1_3$, in which the signal content is expected to be low and the background rate is substantial, dominated by events with two genuine muons from \cPqb\ decays and a non-muon track misidentified as a muon. We do not use data in this topology to search for signal; these events are instead used as a control region to test background parameterizations.

The background rates in the selected topologies are expected to be low with the exception of topology $R^1_{2}$, where backgrounds remain non-negligible even after the $p_T^{\mu\mu}>80$~GeV/$c$ requirement. However, because the search for new resonances is performed in small windows in the invariant mass distribution of muon pairs, the rate of remaining background in each window is comparable to the rate of the signal being sought. For topology $R^1_2$, the main SM background contributions are due to \cPqb\cPaqb\ production with one of the \cPqb\ quarks undergoing a double semileptonic decay, low-mass resonance production (prompt or from heavy-flavour decays), low-mass Drell-Yan production, and occasional muon misidentifications due to decays-in-flight, either alone or in combination with a muon from a heavy-flavour decay.  For topology $R^2_{22}$, the SM backgrounds are dominated by \cPqb\cPaqb\ production with both \cPqb\ jets undergoing double semileptonic decays or fragmenting into low-mass resonances decaying to pairs of muons. Background events with muon jets consisting of multiple muon candidates (the quadmuon topology $R^1_{4}$ and the higher-order topologies) typically originate from events with two muons from a \cPqb\ jet and the other muons from either decays-in-flight, punch-through, or muon misidentifications where some of the segments from true muons are matched to the non-muon tracks. The SM content of the higher order regions is due to rare combinations of the mechanisms discussed above and is extremely low.

\begin{figure}[tbh]
\centering
\begin{tabular}{cc}
\includegraphics[width=0.45\linewidth]{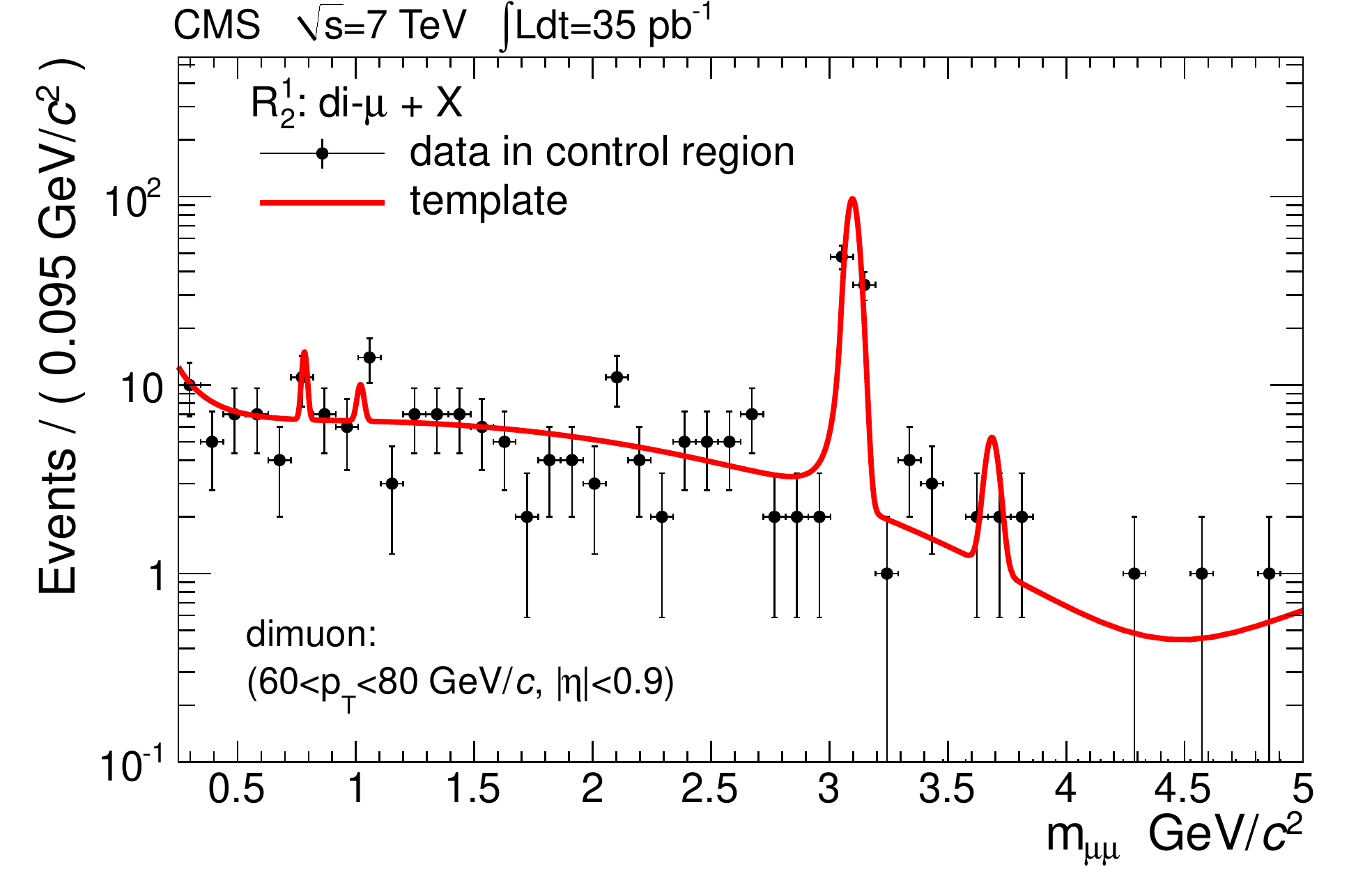}      \put(-40,100){\large (a)}&
\includegraphics[width=0.45\linewidth]{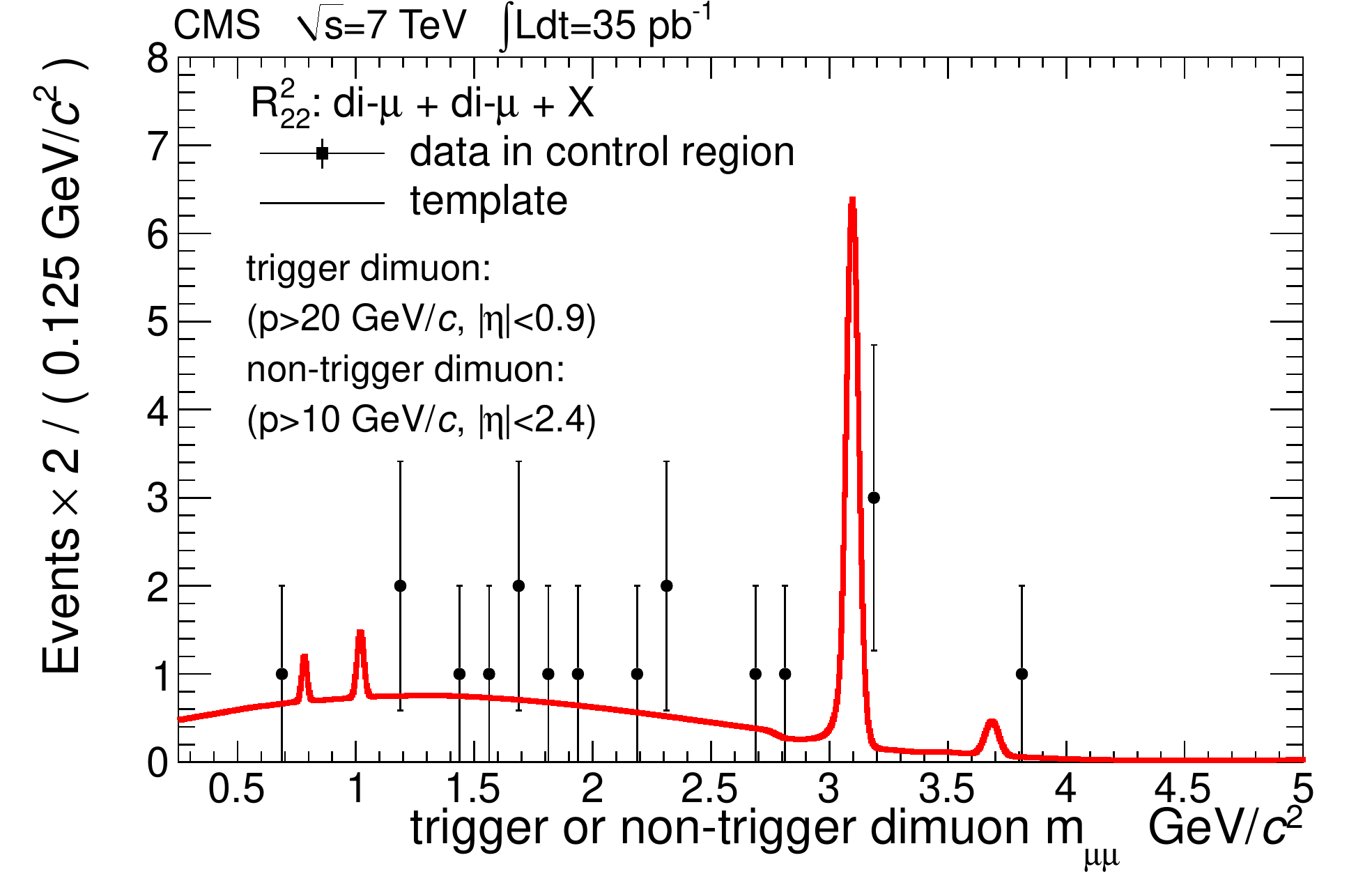}     \put(-40,100){\large (b)}\\
\includegraphics[width=0.45\linewidth]{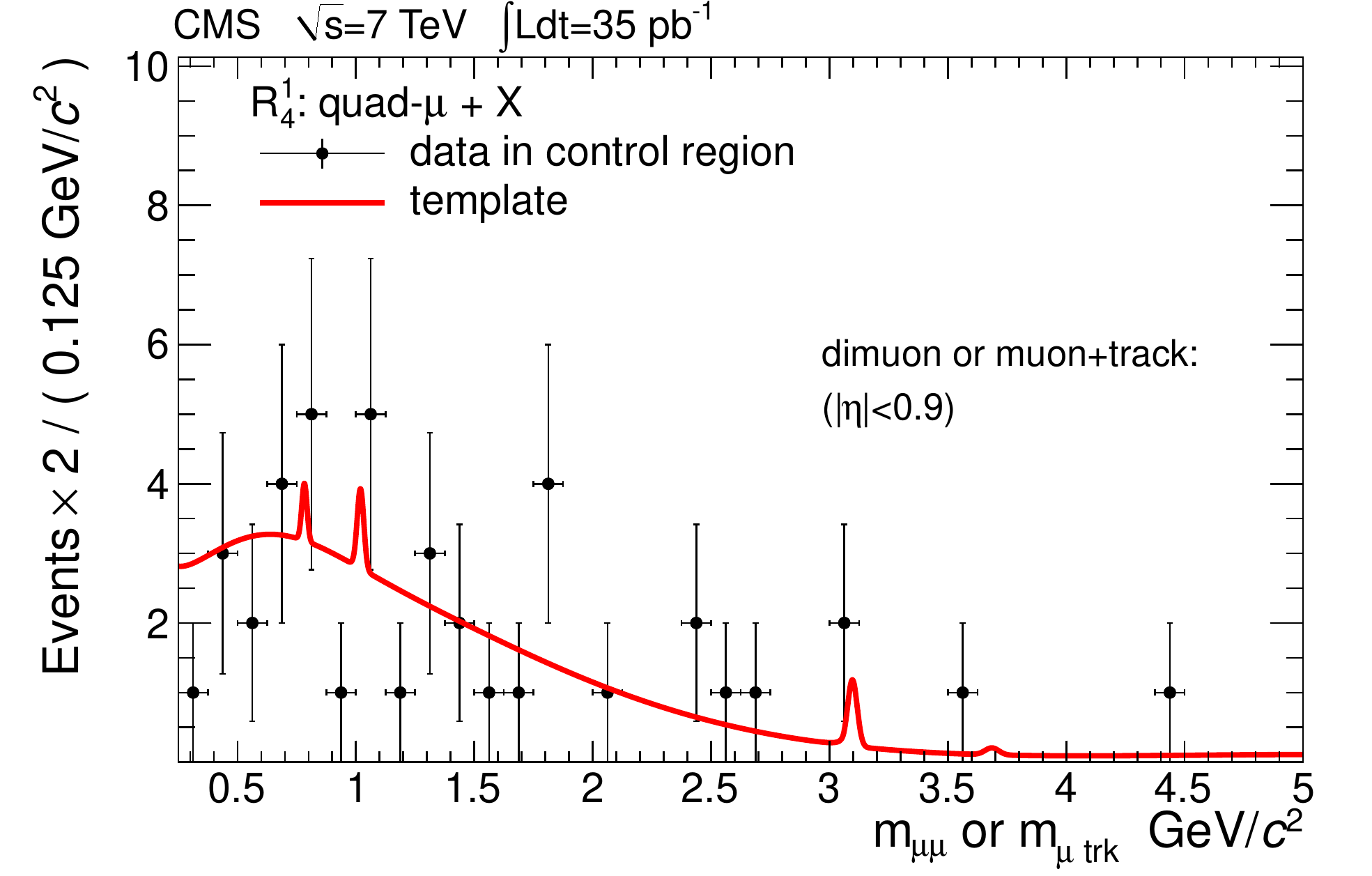}     \put(-40,100){\large (c)} &
\includegraphics[width=0.45\linewidth]{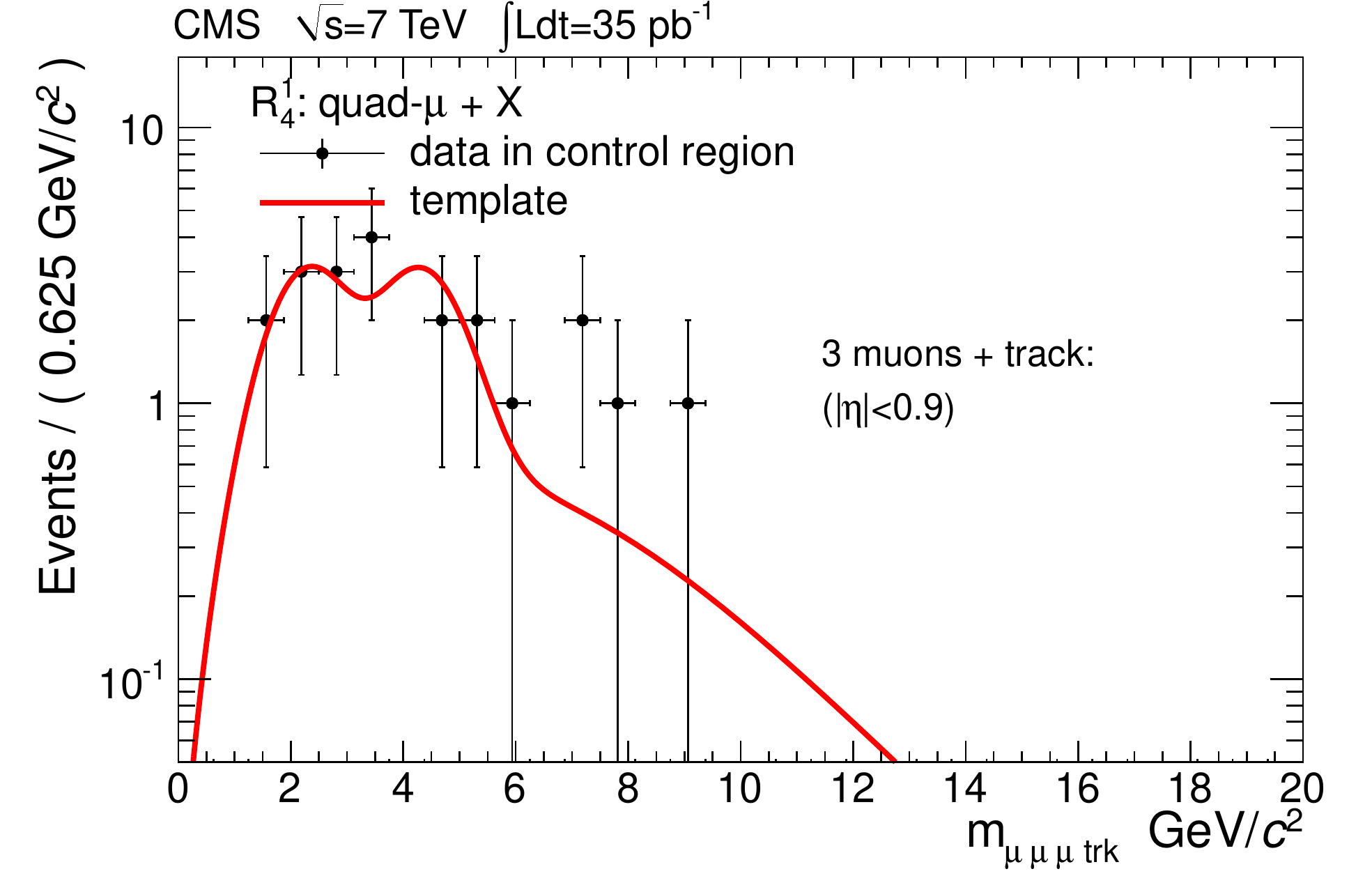}   \put(-40,100){\large (d)} \\
\end{tabular}
\caption{
(a): Data in the single-dimuon category ($R^1_2$) control region $60<p_T^{\mu\mu}<80$~GeV/$c$ overlaid with the background prediction obtained from the background-enriched region $40<p_T^{\mu\mu}<60$~GeV/$c$, fitted for overall normalization only.
(b): The invariant mass of all dimuons in the off-diagonal region for events in the two-dimuon category ($R^2_{22}$; note that there are two entries per event), compared with the prediction obtained from the full 2D background template, fitted for overall normalization only.
(c): The invariant mass distribution of all ``dimuons'' in the $3\mu+$track events (two entries per event) used as a control region for the analysis of events in the quadmuon category ($R^1_4$). The distribution is compared with the prediction obtained from the full 2D background template, fitted for overall normalization only.
(d): The invariant mass of the four ``muons'' in the $3\mu+$track control region for the quadmuon topology $R^1_4$ compared with the prediction obtained from the data in the background-enriched region ($2\mu+2$tracks).
\label{fig:control_distributions}}
\end{figure}

To account for background contributions, we construct ``templates'' (one for each topology) modeling the distribution of reconstructed muon pair masses in background events. With the exception of the single dimuon topology $R^1_2$, the templates are multidimensional distributions in the $(m_1, ..., m_K)$ space of reconstructed muon pair masses, where $K$ is the number of dimuons characteristic of a given topology. For each category, we define one or more background-enriched regions used to construct the template. In addition, we define a control region for validating the template using events with properties closely resembling those of the final events or using the off-diagonal side-band of the final $(m_1, ..., m_K)$ distribution.

While the templates were derived directly from data, we use simulation to determine the composition of the backgrounds as well as momentum evolution of certain parameters, e.g.\ the dimuon mass resolution and shape of the invariant mass distributions. To ensure that simulation is reliable in the phase space characteristic of this analysis, a series of detailed studies have been performed. First, the single-dimuon dataset with $p_T^{\mu \mu}<80$~GeV/$c$ (low momentum part of topology $R^1_2$) was divided in subsets, each one dominated by only one of the contributing background processes to measure rates, shapes, kinematic distributions, tracking related parameters and resolutions, and mass resolutions of low-mass resonances as a function of the boost. These measurements were compared to simulation predictions and showed very good agreement except for a few known and well-understood shortcomings of the simulated samples available (lack of very low-mass Drell-Yan events, missing production modes and/or decay channels for some resonances). In the following, we describe the details of the procedure to construct the background templates for topologies $R^1_2$, $R^2_{22}$ and $R^1_4$, which have the highest background content. Templates for higher order topologies are constructed in a similar fashion, but ultimately have not been used explicitly in the final fit because no data events were observed in topologies with more than four muons (the Bayesian limit is insensitive to the rate of expected background events when no data events are observed). These topologies still contribute to the analysis sensitivity as benchmark models predict a non-negligible fraction of signal acceptance to fall into the higher order topologies.

To model the shape of the invariant mass distribution for the single dimuon region $R^1_2$, we define two sub-regions with $40<p_T^{\mu \mu}<60$~GeV/$c$ (background-enriched region) and $60<p_T^{\mu \mu}<80$~GeV/$c$ (control region). The first sub-region is used to obtain a parametrization of the shape of the dimuon invariant mass distribution in background events. The data were parameterized using a combination of Bernstein polynomials~\cite{Bernstein} and Crystall Ball functions~\cite{crystal-ball}  describing resonances, and the results are shown in Fig.~\ref{fig:background_distributions}(a). To validate the template with data, we fit its shape to the observed data in the region $60<p_T^{\mu \mu}<80$~GeV/$c$, allowing only the overall normalization to float in the fit. To account for the evolution of the resolution of mass measurement with $p_T^{\mu\mu}$, an additional term was added to the uncertainties in the widths of resonances, and we used simulation to verify that the shape of the bulk of the background distribution is only a weak function of $p_T^{\mu\mu}$. The comparison shows good agreement as illustrated in Fig.~\ref{fig:control_distributions}(a). The same template (with the uncertainty of mass resolution evolved to cover even higher $p_T^{\mu\mu}$) is used to predict the shape of background events of topology $R^1_2$ in the signal region $p_T^{\mu\mu}>80$~GeV/$c$.

The SM backgrounds in the two-dimuon topology $R^2_{22}$ are dominated by \cPqb\cPaqb\ events with each \cPqb\ quark yielding a pair of muons. Because each \cPqb\ jet fragments independently, the background distribution in the $(m_1,m_2)$ space of the two dimuon masses is a Cartesian product of the 1D dimuon mass distribution with itself. However, because one of the dimuons contains the $p_T > 15$~GeV/$c$ muon that triggered the event, its dimuon mass spectrum is different from that of the other dimuon. To account for this effect, we separately measure the shapes for the ``trigger'' and ``non-trigger'' dimuons. To model the trigger dimuon shape, we use single-dimuon events with further selections suppressing the non-\cPqb\cPaqb\ backgrounds, fit to a parameterized functional form and subtract the residual contamination from Drell-Yan background, which does not contribute to the topology $R^2_{22}$ (both subtracted and non-subtracted curves are shown in Fig.~\ref{fig:background_distributions}(b)). To match the kinematics of the two-dimuon events being modeled, the ``other dimuon'' shape is obtained using three-muon events with a dimuon recoiling off a trigger-quality muon and is shown in Fig.~\ref{fig:background_distributions}(c). To account properly for a contribution with both dimuons containing a trigger-quality muon in the final 2D template, an additional reweighing is applied in taking the Cartesian product of the two distributions. The template is validated using finally selected two-dimuon events in the off-diagonal part of the $(m_1,m_2)$ distribution. Figure~\ref{fig:control_distributions}(b) shows a comparison of the invariant mass distribution of the dimuons in these events (note that there are two histogram entries per event), compared to the prediction derived from the template and fit to the data for overall normalization only.

The quadmuon topology $R^1_4$ has a small background contamination in which a \cPqb\ quark produces two genuine muons and additional two muons are produced from non-muon tracks incorrectly matched to some of the genuine muon stubs. When identifying the two fundamental dimuons within the group of four muons, both $(\mu , \mu)+(trk ,trk)$ and $(\mu ,trk)+ (\mu, trk)$ pairings can occur, each having its own distinct 2D shape in the $(m_1,m_2)$ space. To model these events, we use single dimuon events and construct ``pseudo muon jets'' using two reconstructed muons and two non-muon tracks playing the role of misidentified muons. Selected events are separated into two subsets according to the type of pairing, each producing a 2D distribution for the invariant masses of the two pairs in the event. Figures~\ref{fig:background_distributions}(d), (e) and (f) show 1D invariant mass distributions for $(\mu , \mu)$, $(trk ,trk)$, and $(\mu ,trk)$-type dimuons obtained from projections of the 2D distributions for the two types of events. In the $R^1_4$ signal events, the identities of $\mu$ and $trk$ are unknown, so the mass templates and the signal events are both symmetrized by randomly assigning dimuon masses to the horizontal and vertical axes of the 2D distribution. The template is validated using a control region with three nearby muon candidates ($R^1_3$), one of which is likely a misidentified hadron, and adding a non-muon track to play the role of a second misidentified muon. Figure~\ref{fig:control_distributions}(c) compares the distribution of all ``dimuons'' in the $3\mu+$track control sample (note two entries per event) compared to the prediction based on the full 2D template fitted to data for overall normalization only. Figure~\ref{fig:control_distributions}(d) makes a similar comparison but for the quadmuon invariant mass. Templates for higher order topologies are derived as combinations of the above methods. In all cases, the full posterior density functions for fit parameters including correlations were used in the final fit to account for the uncertainties in the background templates.

\begin{figure}[tbh]
\centering
\begin{tabular}{cc}
\includegraphics[width=0.4\linewidth]{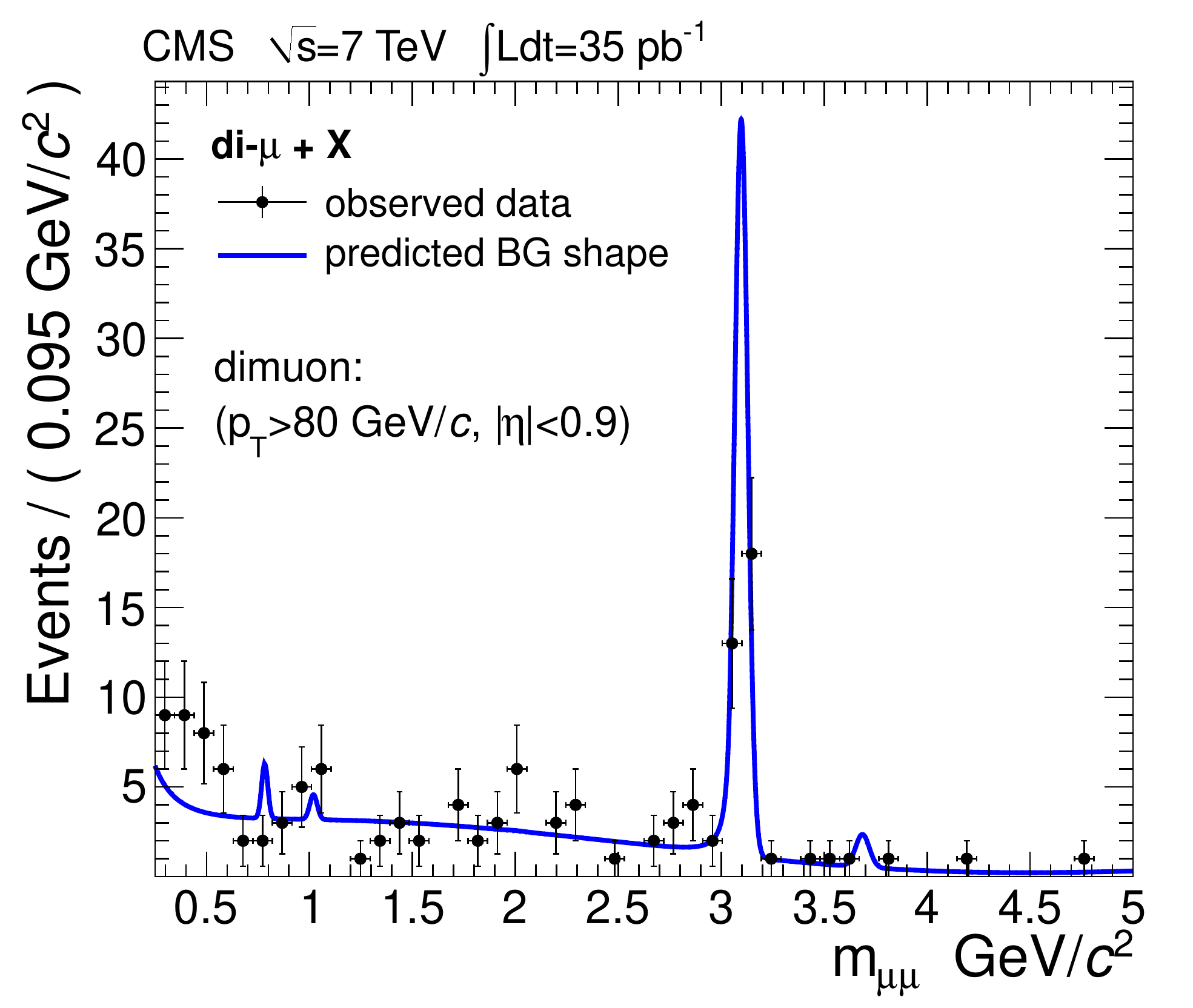}      \put(-40,30){\large (a)} &
\includegraphics[width=0.4\linewidth]{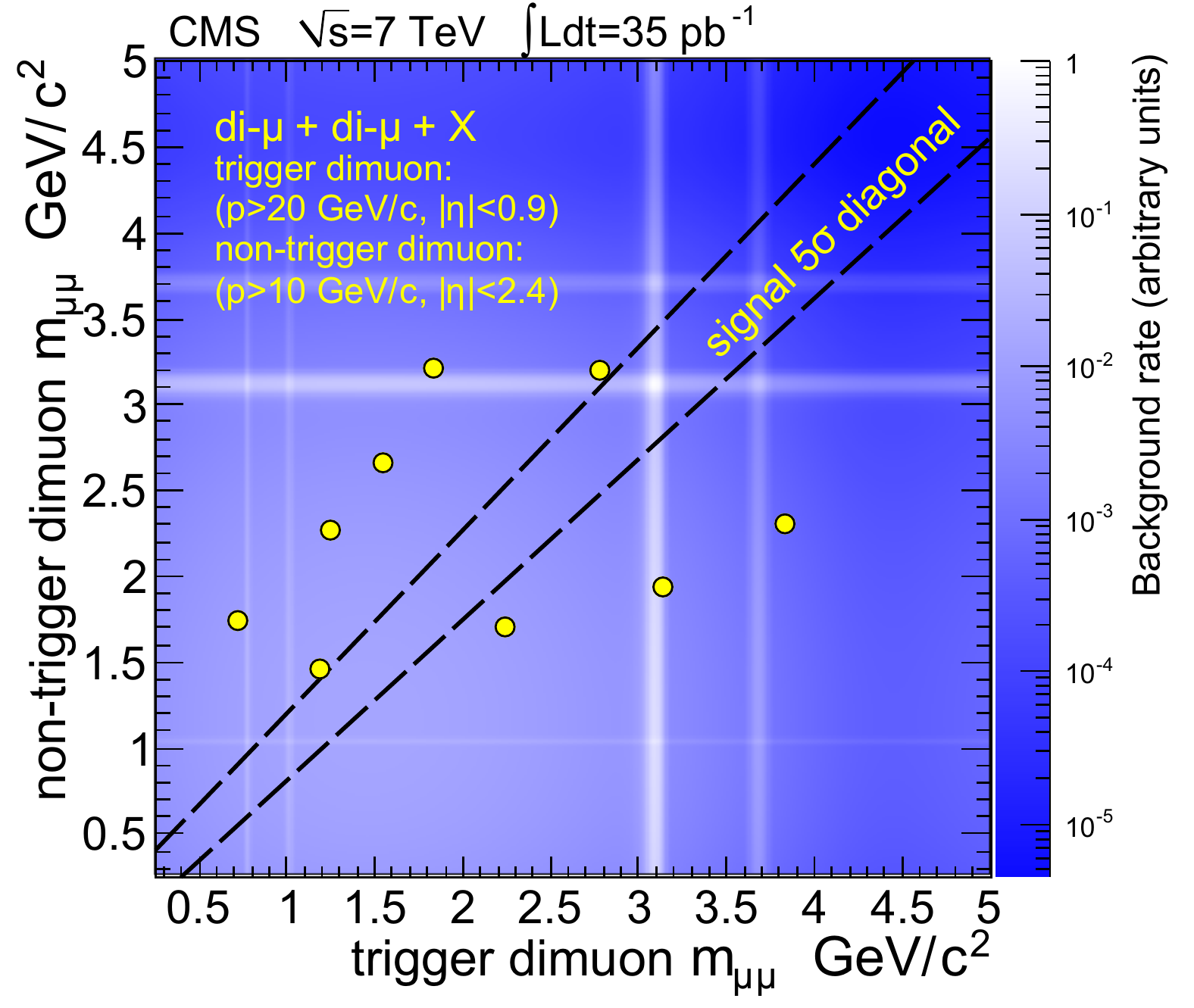}      \put(-50,30){\large (b)}\\
\includegraphics[width=0.4\linewidth]{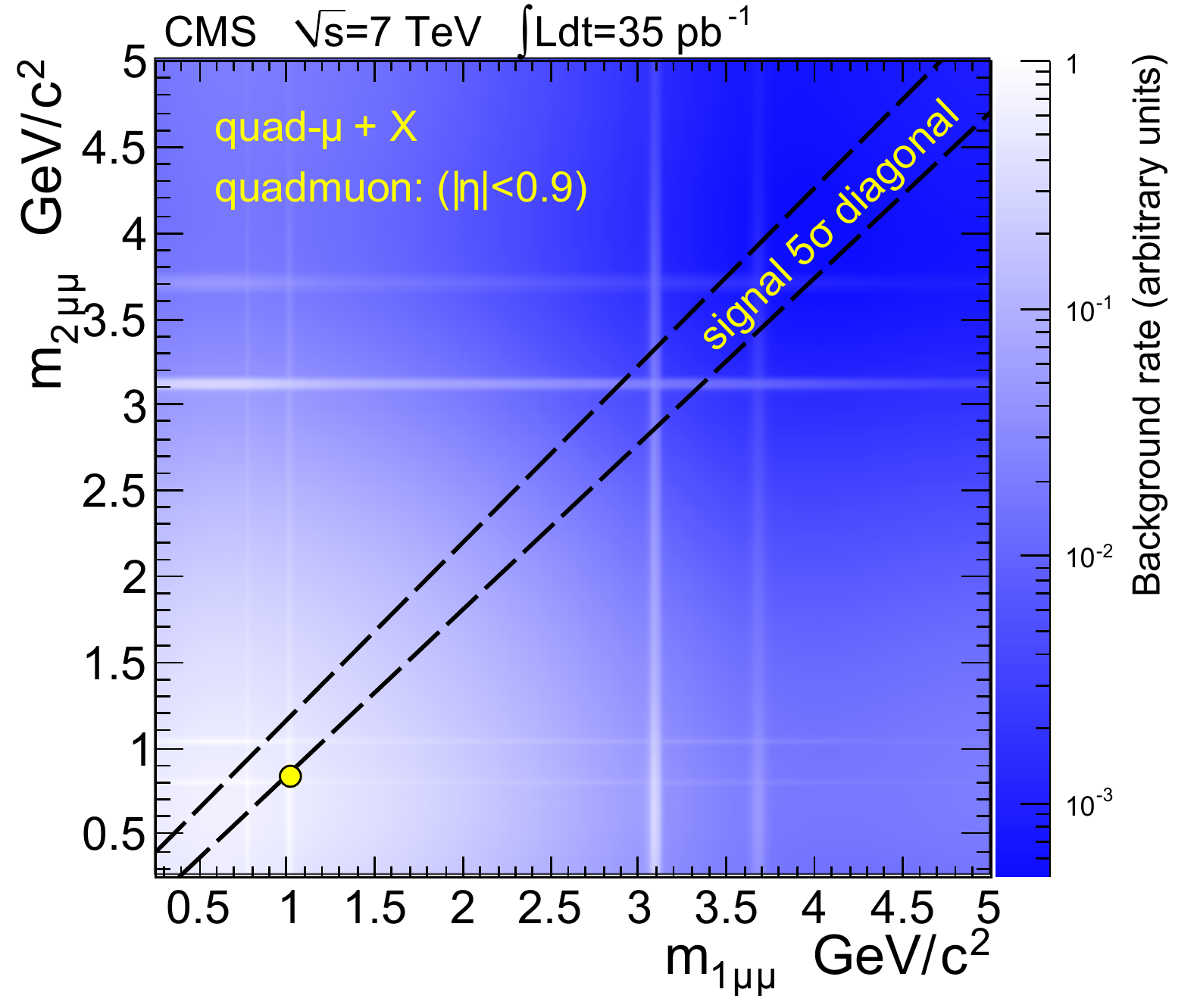}     \put(-50,30){\large (c)} &
\includegraphics[width=0.4\linewidth]{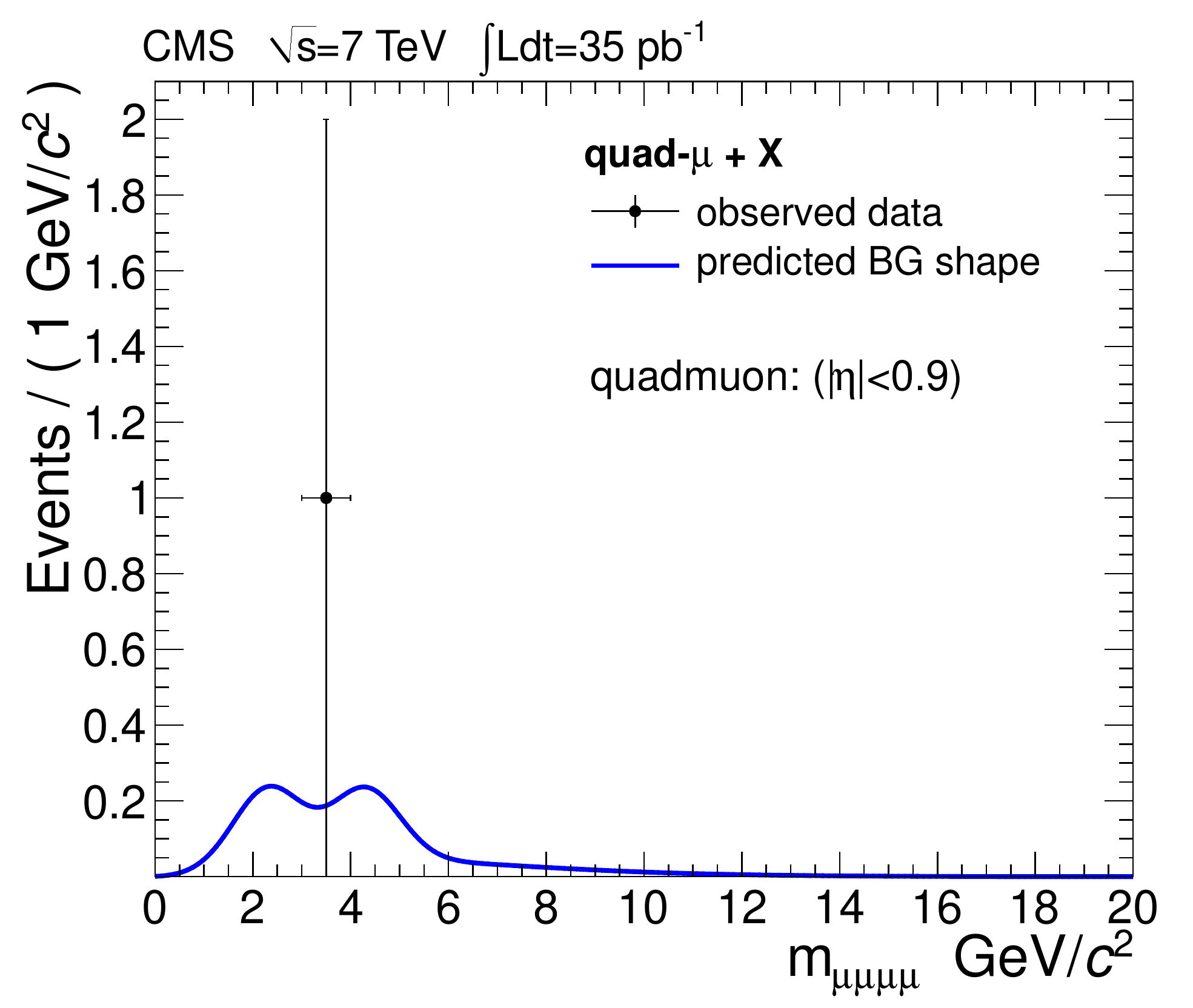}     \put(-40,30){\large (d)} \\
\end{tabular}
\caption{The 1D and 2D invariant mass distributions of muon pairs for events in each signal region, compared with the expected background. (a): Events in the single dimuon topology $R^1_2$. (b): 10 events in the two-dimuon topology $R^2_{22}$. (c): The single ``quadmuon'' event in topology $R^1_4$. (d): The invariant mass of all four muons for the same event. None of the events in the multi-dimensional topologies fall into the corridor along the diagonal (shown as dashed lines), which would indicate the presence of signal. The last plot is relevant for the special scenario with a cascade decay $a_2 \to a_1 a_1$ with $m(a_2)<2 m(a_1)$, leading to the off-shell production of $a_1$. \label{fig:signal_distributions}}
\end{figure}

\subsection{Efficiencies and Systematic Uncertainties}
The shape of the invariant mass distribution for possible signal events was studied by comparing the properties of events with dimuons from $\omega$, $\phi$, $\JPsi$, and $\psi^{\prime}$ resonances in data with the simulation predictions and extrapolating between the resonance masses. Because of the excellent resolution of the CMS tracker, signal shapes have narrow widths scaling with the mass of the resonances, with a slight dependence on the dimuon transverse momentum. For final fits, signal shape is parameterized using a Crystal Ball function with core resolution of $\sigma(m_{\mu\mu}) = 0.026\;{\rm{GeV}}/c^2  + 0.0065\, m_{\mu\mu} $ for barrel ($|\eta^{\mu\mu}| < 0.9$) and $0.026\;{\rm{GeV}}/c^2  + 0.013\, m_{\mu\mu} $ for endcap ($0.9 < |\eta^{\mu\mu}| < 2.4$). To account for the unknown momentum range of the hypothetical signal, the width of the Gaussian core was allowed to vary within 30\%. Multidimensional distributions were obtained by taking appropriate Cartesian products. The uncertainties on the parameters of the function are obtained by quantifying the level of agreement between data and simulation, which is dominated by the statistical uncertainties.
For dimuons with $p_T^{\mu\mu}<150$~GeV/$c$, the reconstruction efficiency in the barrel region is nearly flat as a function of $\eta$, has an average value of $95 \pm 1$\%, and is driven by the efficiency in finding and matching stubs in the muon system. The efficiency decreases to about 92\% for $m_{\mu\mu}$ close to $2m_\mu$ because of muon trajectories becoming nearly collinear. In the endcap region, there is a slight lowering of efficiency towards high $|\eta|$ because of muon trajectories overlapping in the muon system.
The systematic uncertainty on the efficiency in the endcap region is 3\% as obtained from comparisons of reconstruction efficiencies for simulated events with overlapping and non-overlapping muons, both with each other and with the data. For higher momentum dimuons, tracking effects become important as the overlaps of the trajectories lead to high sharing of hits and decreased efficiency. This effect is especially pronounced for $m_{\mu\mu}$ of 0.4--0.6 GeV/$c^2$, where the efficiency decreases to $85\pm 5$\% at $p_T^{\mu\mu}\approx 250$~GeV/$c$ and to $75\pm 10$\% at $p_T^{\mu\mu}\sim 350$~GeV/$c$ because of events in which muon trajectories bend towards each other in the magnetic field and remain close to each other for a substantial part of their paths through the tracker. For muon pairs with $m_{\mu\mu}$ outside of this range, the efficiency is nearly flat until $p_T^{\mu\mu}\approx250$~GeV/$c$, where it starts a slow descent as $p_T^{\mu\mu}$ increases; it is about $85\pm5$\% at $p_T^{\mu\mu}\approx 350$~GeV/$c$. The uncertainties quoted account for possible inaccuracies in modelling the size of the tracker hit clusters in the silicon tracker potentially affecting the rate of merging of hits from nearby tracks, which is the dominant cause of the inefficiency of track reconstruction.

Reconstruction of quadmuons suffers more significantly from reconstruction failures in the muon system. In addition to the 3\% inefficiency per muon, the probability to reconstruct all four muons in a quadmuon has a significant additional term related to small uninstrumented gaps between the wheels in the central part of the muon system. With an inefficiency of 8--10\% per muon crossing the gap region and a significant probability for one or more muons to cross it, the average reconstruction efficiency for a quadmuon is $83\pm 3$\% ($p_T^{\mu\mu\mu\mu}\approx 150$ GeV/$c$). For high-momentum quadmuons, tracking effects become more significant. In topologies with on-shell dimuons relevant to this analysis, the quadmuon efficiency is determined by the probability to reconstruct separately each of the two low-mass dimuons comprising the quadmuon, as significant overlaps of trajectories for muons from different dimuons are rare. The momentum dependence of the quadmuon efficiency closely follows the reconstruction efficiency of a dimuon with momentum equal to the higher-momentum dimuon within the quadmuon. The reconstruction efficiency also has the same dependence on the invariant mass of the dimuons within the quadmuon. For a quadmuon with constituent dimuon mass of about 0.5~GeV/$c^2$ (the worst-case scenario), the average efficiency at $p_T^{\mu\mu\mu\mu}\sim250$~GeV/$c$ is about 74\% and has a systematic uncertainty of 2--3\% due to the tracking efficiency. Higher-multiplicity muon jets have larger inefficiencies because of muon system reconstruction failures, but are reconstructed as lower-multiplicity jets. In the context of a specific model, the migration of events between the high-multiplicity topologies does not reduce the overall acceptance. Higher multiplicity muon jet reconstruction is less affected by tracking inefficiencies as the average momentum of dimuons is moderate, even for very high momentum ($\sim400$~GeV/$c$) muon jets.

\section{Results}

The data in the regions used to search for signal (the ``diagonal'' regions of multi-dimensional distributions and the single-dimuon events with $p_T^{\mu\mu}>80$ GeV/$c$ in topology $R^1_2$) were not looked at until all analysis selections and signal extraction techniques were finalized. When the signal regions were examined, no evidence for new resonance production was found within the sensitivity of this analysis. Figure~\ref{fig:signal_distributions} shows the observed data for select topologies and the expected SM background contributions, which were obtained using the templates for each topology. The templates were normalized to the data in the off-diagonal regions in high-multiplicity topologies (all but $R^1_2$) and directly fitted in a combined signal-plus-background fit for the single-dimuon case ($R^1_2$).  For topology $R^2_22$, 10 events were observed in the non-diagonal region of the distribution with one event falling outside the $m_{\mu\mu}>5$ GeV/$c$ range and therefore not seen in Figure~\ref{fig:signal_distributions}(b). The single off-diagonal event in the quadmuon topology $R^1_4$ can be interpreted in the context of models with new light bosons produced off-shell via a process of the type $a_2 \to a_1 a_1 \to 4\mu$. No events were found in any topologies with more than four muons.

\begin{figure}[tbh]
\centering
\begin{tabular}{cc}
\includegraphics[width=0.45\linewidth]{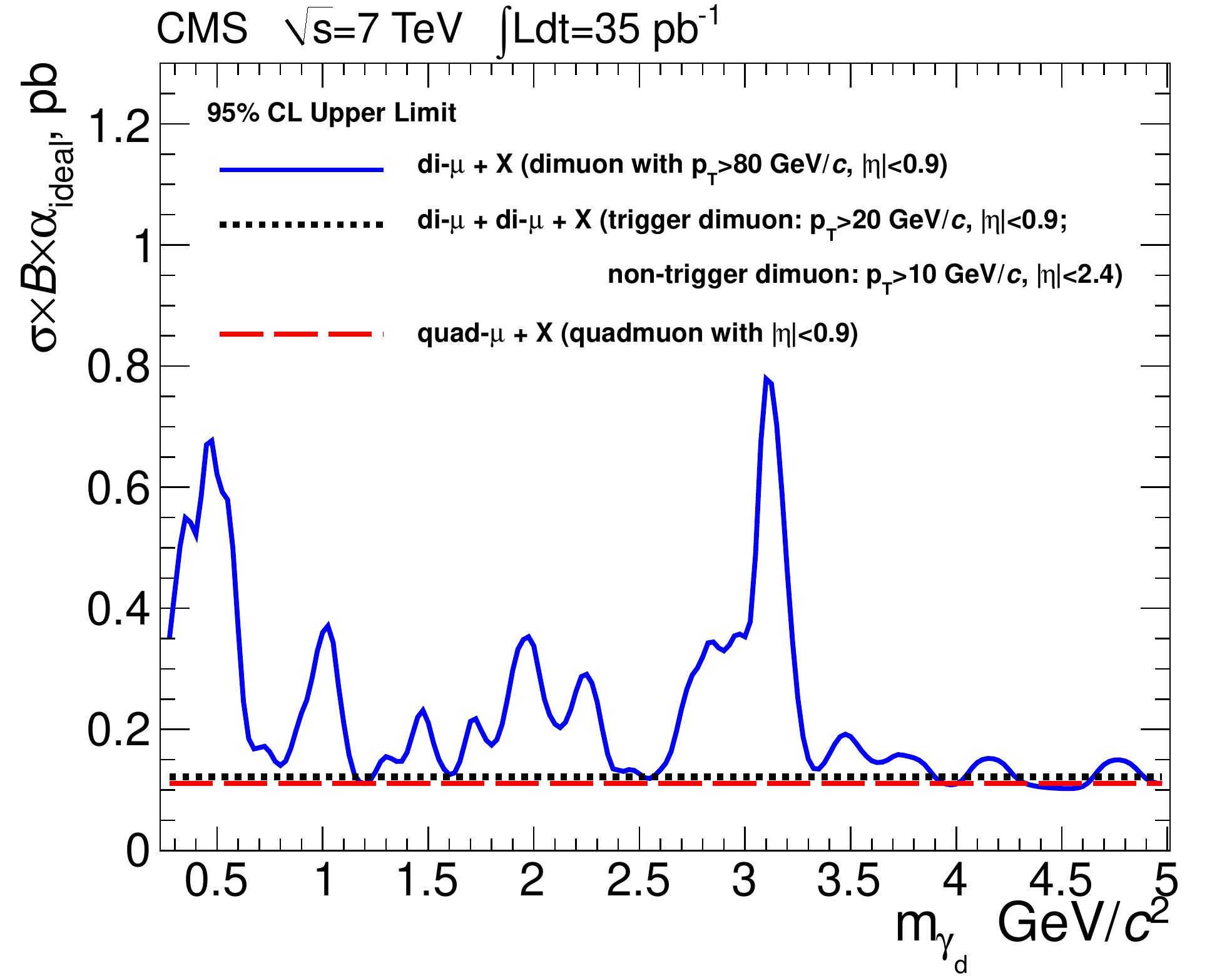}      \put(-150,80){\large (a)} &
\includegraphics[width=0.45\linewidth]{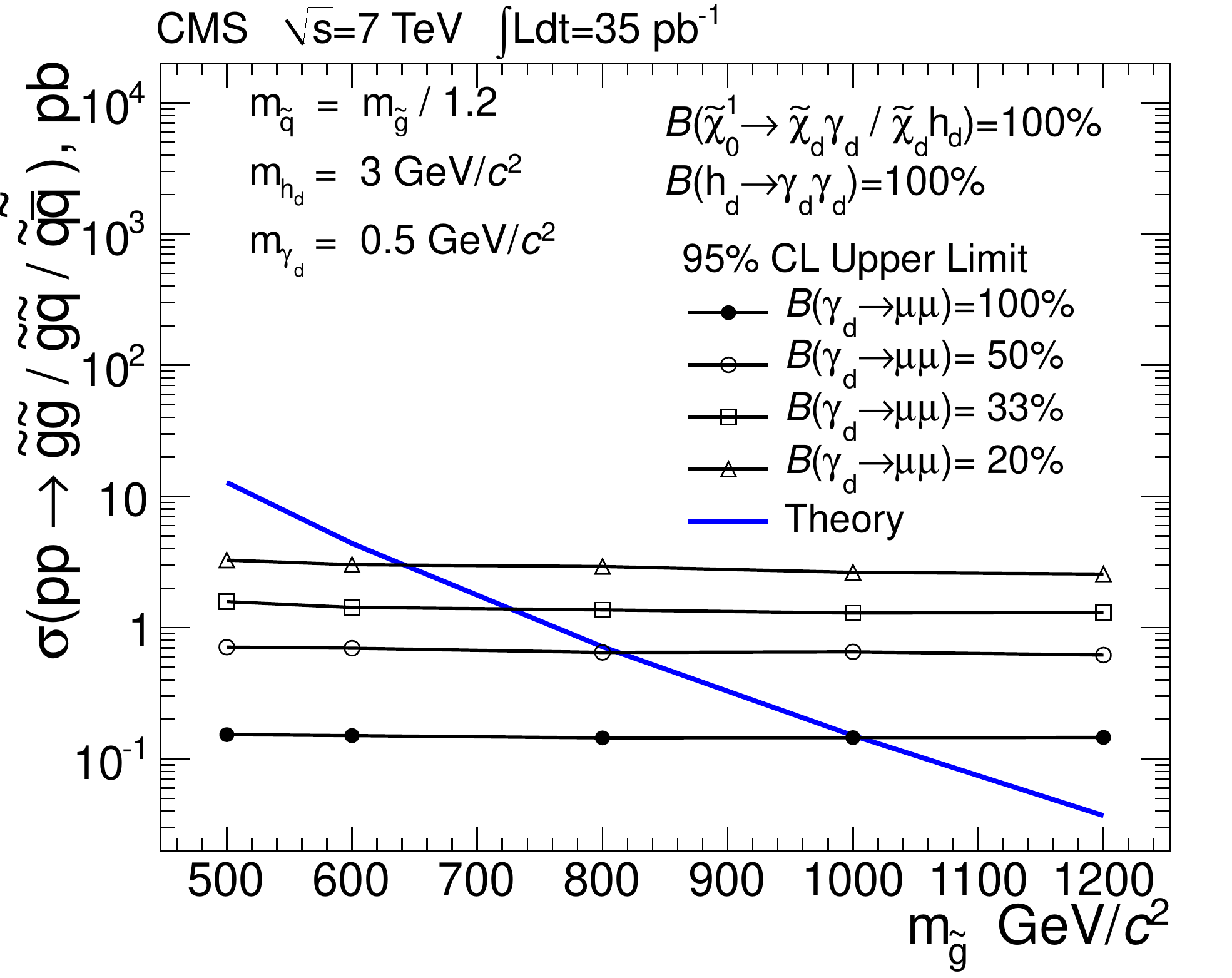}     \put(-150,80){\large (b)} \\
\includegraphics[width=0.45\linewidth]{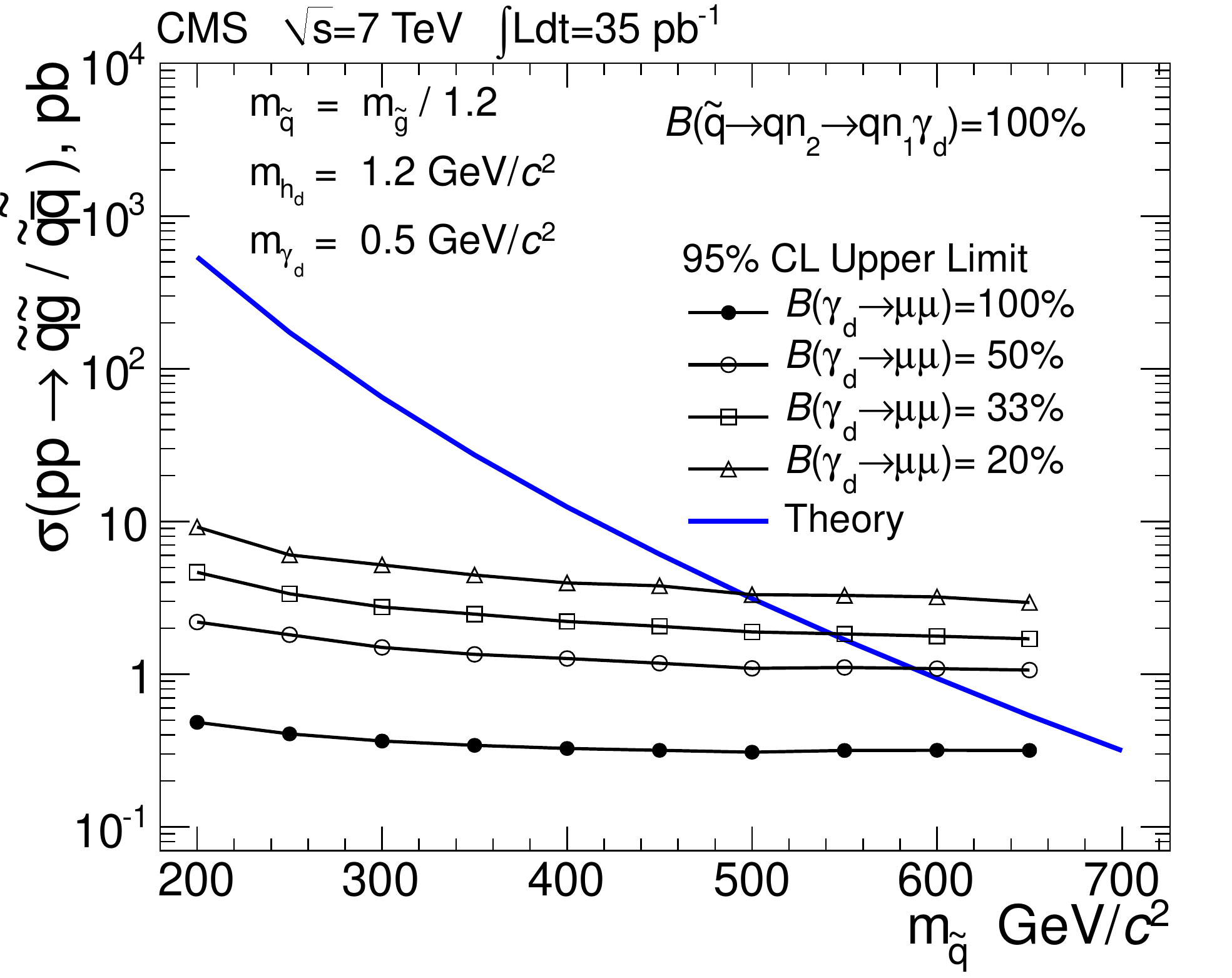}      \put(-150,80){\large (c)} &
\includegraphics[width=0.45\linewidth]{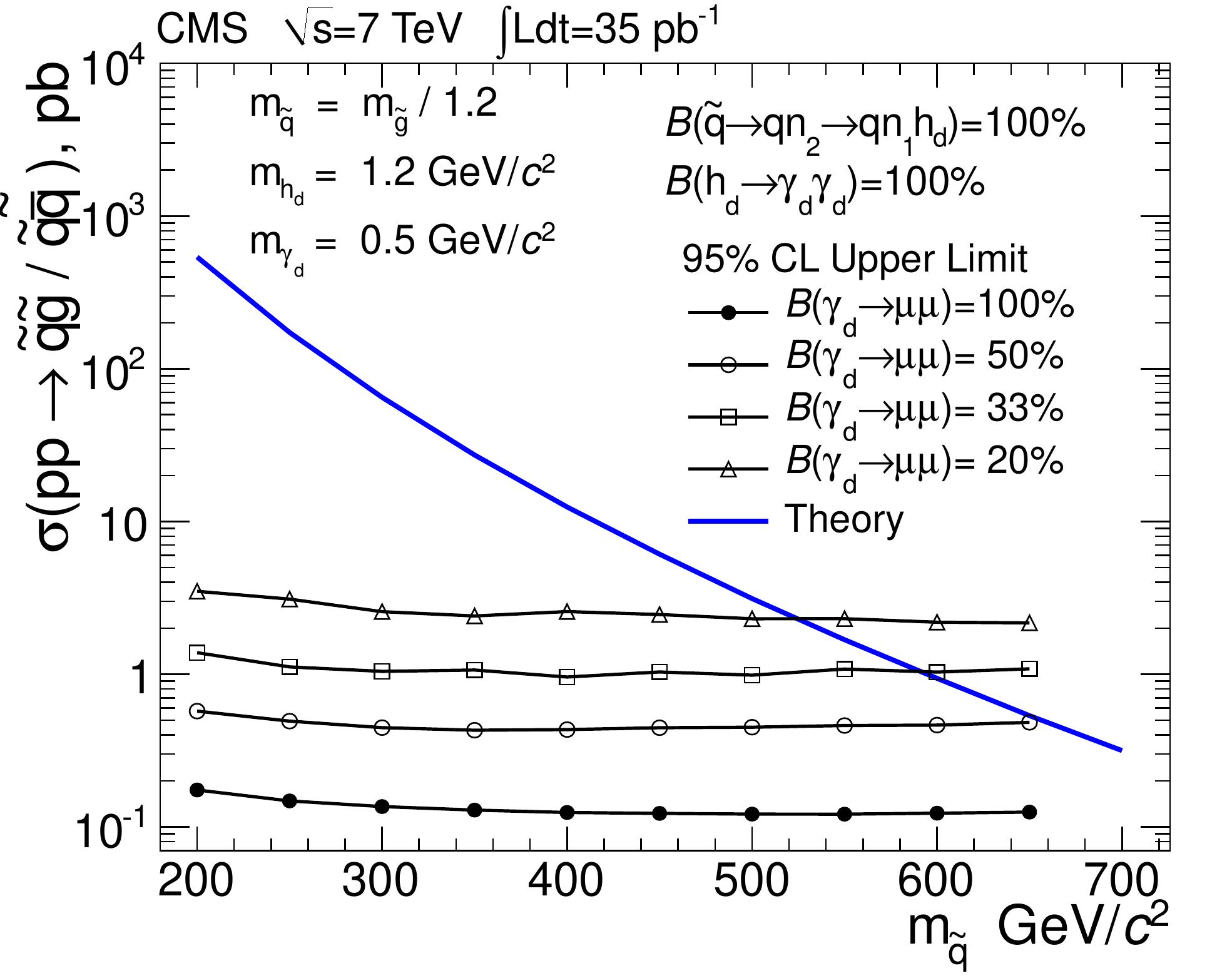}     \put(-150,80){\large (d)} \\
\end{tabular}
\caption{(a): 95\% CL upper limits on the rate of the signals of new physics with leptonic jets for three topologies expressed as cross section times appropriate branching fractions to produce a particular signature (including branching fraction for $\gamma_{\rm dark} \to \mu \mu$) times kinematic and geometrical acceptance (muon momentum thresholds and $\eta$ ranges), assuming an ideal detector. (b): Limits for the Dark SUSY model~\cite{BaiHan} with the MSSM LSP decaying via $\tilde{\chi}^0_1 \to \tilde{\chi}_{\rm dark} \gamma_{\rm dark}+\tilde{\chi}_{\rm dark} h_{\rm dark} (\to \gamma_{\rm dark} \gamma_{\rm dark})$, with the $\tilde{\chi}_{\rm dark}$ being the new LSP. (c) and (d): Limits on the model~\cite{Ruderman} where squark is the MSSM LSP decaying into a quark and a light hidden sector fermion decaying to a lighter hidden sector fermion with emission of either a dark photon (c) or a light dark-Higgs (d) decaying to two dark photons.
\label{fig:final_limits}}
\end{figure}

To interpret the results in a model-independent fashion, we set the 95\% confidence level (CL) upper limits on the allowed production rate of the single-dimuon+X, quadmuon+X and two-dimuons+X topologies. To simplify further phenomenological interpretations of these results, the rate is defined as the cross section times appropriate branching fractions to produce a particular signature times kinematic and geometrical acceptance, assuming an ideal detector. The ``ideal detector acceptance'' $\alpha_{\mbox{\scriptsize ideal}}$ is the probability for an event to satisfy analysis selections for a specific topology taking into account momentum thresholds and $\eta$ ranges of muons, but ignoring all detector and reconstruction algorithm inefficiencies. For a given model, $\alpha_{\mbox{\scriptsize ideal}}$ can be evaluated using one of the standard event generators and requires no knowledge of the instrumental inefficiencies, which are absorbed into the experimental limit calculation.The upper limits are shown in Fig.~\ref{fig:final_limits}(a) and the main acceptance cuts, defining the ideal detector acceptance for the three topologies, are given in the legend of the same figure.

Limits are set using a Bayesian technique including integration over the systematic uncertainties in the signal and background shape parameterizations and the background normalization, which are treated as nuisance parameters using the true posterior density for the background normalization and log-normal priors in other cases. The limits shown in Fig.~\ref{fig:final_limits}(a) account for systematic uncertainties in the knowledge of instrumental inefficiencies as well as the variations of the reconstruction efficiencies for dimuon masses ranging from $0.25$ to 5~GeV/$c^2$ and the means of the muon jet (dimuon or quadmuon) momentum distributions up to 250~GeV/$c$. These variations were treated as an additional systematic uncertainty and were 7\% for the quadmuon topology, 20\% for dimuon, and 35\% for the two-dimuon topology. For higher momentum muon jets, the limits become weaker because of diminishing track reconstruction efficiency. Other systematic uncertainties account for the precision in the luminosity measurement (4\%), uncertainties in the efficiency of reconstructing and matching muon stubs to the tracks and triggering (1-4\% depending on the topology), and track reconstruction efficiency (5-10\%).

The limits presented can be used as conservative bounds on higher multiplicity topologies, e.g. the limit for the quadmuon+X topology can be used to constrain the production rate of signals with muon jets containing four or more muons in it. These limits are conservative in assuming that reconstruction failures always remove events from consideration, whereas in reality events in which there are reconstruction failures usually enter another, lower multiplicity, topological category. The limit for the quadmuon+X topology can be reinterpreted for models predicting $a_2 \to a_1 a_1 \to 4\mu$ cascades with one or both $a_1$ bosons off-shell (if $m(a_2)<2 m(a_1)$). In this case, the 95\% CL upper limit on the production rate is the same as the quadmuon+X limit in Fig.~\ref{fig:final_limits}(a) except near $m(a_2)\approx 3.5$ GeV/c$^2$, the invariant mass of all four muons in the single off-diagonal event observed in topology $R^1_4$. In that region, the limit can be conservatively estimated by multiplying the quadmuon+X limit by factor of 4.74/3.00 (the ratio of Bayesian 95\% CL upper limits for the mean of a Poisson variable given one and zero observed events in the absence of backgrounds).

To set limits for representative benchmark scenarios, we use two models of SUSY with the dark sector. The first SUSY model~\cite{BaiHan} assumes standard MSSM squark/gluino production and cascades followed by the decay of the MSSM LSP into the dark sector $\chi_1^0 \to h_{\mbox{\scriptsize dark}} \chi_{\mbox{\scriptsize dark}}$ or $\chi_1^0 \to \gamma_{\mbox{\scriptsize dark}} \chi_{\mbox{\scriptsize dark}}$, where $\chi_{\mbox{\scriptsize dark}}$ is the new cold dark matter candidate. The lightest MSSM neutralino mass is set to 400 GeV/$c^2$ and the dark sector masses used are $m(h_{\mbox{\scriptsize dark}})=3$, $m(\gamma_{\mbox{\scriptsize dark}})=0.5$, and $m(\chi_{\mbox{\scriptsize dark}})=300$~GeV/$c^2$. The resulting limits are shown as a function of gluino mass $m(\tilde{g})$ ($m(\tilde{q})=m(\tilde{g})/1.2$ for the first two generations) in Fig.~\ref{fig:final_limits}(b) for three different choices of branching fractions $B(\gamma_{\mbox{\scriptsize dark}} \to \mu \mu)$. The systematic uncertainties used are the same as for model-independent limits with two exceptions. First, we exclude uncertainties used to compensate for variations in efficiency with $p_T^{\mu\mu}$ because in this case the momentum spectra are known. Second, we include an uncertainty of 3\% in the acceptance to account for uncertainties in proton parton distribution functions (PDF) by varying parameterizations within the CTEQ6.6~\cite{cteq6} family, and comparing the central values of CTEQ6.6L with  NNPDF2.0~\cite{nnpdf}, and MSTW2008~\cite{mstw} sets. As a reference, typical signal acceptance (the probability for an event to be reconstructed in one of the signal topologies) is of the order of 70--75\% for $m(\tilde{q})=$500--1000 GeV/c$^2$ assuming $B(\gamma_{\mbox{\scriptsize dark}} \to \mu \mu)=100$\%. For $B(\gamma_{\mbox{\scriptsize dark}} \to \mu \mu)=50$\% and 33\%, the corresponding ranges change to 30--40\% and 20-25\%, respectively.

The second model~\cite{Ruderman} assumes squarks to be the MSSM LSP ($m(\tilde{q})=m(\tilde{g})/1.2$). Following production, squarks decay via $\tilde{q} \to q n_2$, where $n_2$ is a heavier dark sector fermion with the decay modes dominated by either $n_2 \to n_1 \gamma_{\mbox{\scriptsize dark}}$ ($n_1$ is a lighter dark fermion) or $n_2 \to n_1 h_{\mbox{\scriptsize dark}}(\to \gamma_{\mbox{\scriptsize dark}}\gamma_{\mbox{\scriptsize dark}})$. For each of the two sub-models the limits on the production cross section are shown in Figs.~\ref{fig:final_limits} (c) and (d) for three different choices of branching fractions $B(\gamma_{\mbox{\scriptsize dark}} \to \mu \mu)$. The dark sector masses are set to $m(h_{\mbox{\scriptsize dark}})=1.2$, $m(h_{\mbox{\scriptsize dark}})=0.5$, $m(n_2)=2$, $m(n_1)=0.5$~GeV/$c^2$. The cross section curves shown in Figs.~\ref{fig:final_limits}(c-d) assume universality of squark masses across three squark generations. The cross section is reduced if squark masses are not universal. The limits presented are the most stringent to date for models with dark SUSY sector from collider experiments. 

%% file: conclusions.tex
\section{Summary}
A topology-based search for groups of collimated muons (muon jets) using a data sample corresponding to an integrated luminosity of 35 pb$^{-1}$ revealed no signal of new physics within the sensitivity of the measurement. No events consistent with two or more decays of the new light boson species to a pair of muons were found in the data, and no excess over the SM backgrounds was observed for production of single high-$p_T$ bosons decaying to pair of muons. Relaxing the assumption that the light bosons are on-shell and searching for possible $a_2 \to a_1 a_1 \to 4\mu$ 
cascades revealed one event, consistent with the background expectation. 

With these observations, we set limits on the production of new low-mass states
decaying to pairs of muons that can be applied to a broad class of models predicting
the leptonic jet signatures. These 95\% CL limits exclude production of new physics in several event 
topologies with $\sigma \times {B} \times \alpha_{\mbox{\scriptsize ideal}}$ in the range of 0.1--0.5~pb, where $\alpha_{\mbox{\scriptsize ideal}}$ refers to the kinematic and geometrical acceptance of the analysis for an 
ideal detector. We also set model-dependent limits on several benchmark models predicting production of the new light states 
in the context of dark SUSY significantly extending the sensitivity of previous searches performed at the Tevatron. 

\section*{Acknowledgements}
We wish to congratulate our colleagues in the CERN accelerator departments for the excellent performance of the LHC machine. We thank the technical and administrative staff at CERN and other CMS institutes, and acknowledge support from: FMSR (Austria); FNRS and FWO (Belgium); CNPq, CAPES, FAPERJ, and FAPESP (Brazil); MES (Bulgaria); CERN; CAS, MoST, and NSFC (China); COLCIENCIAS (Colombia); MSES (Croatia); RPF (Cyprus); Academy of Sciences and NICPB (Estonia); Academy of Finland, MEC, and HIP (Finland); CEA and CNRS/IN2P3 (France); BMBF, DFG, and HGF (Germany); GSRT (Greece); OTKA and NKTH (Hungary); DAE and DST (India); IPM (Iran); SFI (Ireland); INFN (Italy); NRF and WCU (Korea); LAS (Lithuania); CINVESTAV, CONACYT, SEP, and UASLP-FAI (Mexico); MSI (New Zealand); PAEC (Pakistan); SCSR (Poland); FCT (Portugal); JINR (Armenia, Belarus, Georgia, Ukraine, Uzbekistan); MST and MAE (Russia); MSTD (Serbia); MICINN and CPAN (Spain); Swiss Funding Agencies (Switzerland); NSC (Taipei); TUBITAK and TAEK (Turkey); STFC (United Kingdom); DOE and NSF (USA).

%% file: EXO-11-013-authorlist.tex
\textbf{Yerevan Physics Institute,  Yerevan,  Armenia}\\*[0pt]
S.~Chatrchyan, V.~Khachatryan, A.M.~Sirunyan, A.~Tumasyan
\vskip\cmsinstskip
\textbf{Institut f\"{u}r Hochenergiephysik der OeAW,  Wien,  Austria}\\*[0pt]
W.~Adam, T.~Bergauer, M.~Dragicevic, J.~Er\"{o}, C.~Fabjan, M.~Friedl, R.~Fr\"{u}hwirth, V.M.~Ghete, J.~Hammer\cmsAuthorMark{1}, S.~H\"{a}nsel, M.~Hoch, N.~H\"{o}rmann, J.~Hrubec, M.~Jeitler, W.~Kiesenhofer, M.~Krammer, D.~Liko, I.~Mikulec, M.~Pernicka, H.~Rohringer, R.~Sch\"{o}fbeck, J.~Strauss, A.~Taurok, F.~Teischinger, P.~Wagner, W.~Waltenberger, G.~Walzel, E.~Widl, C.-E.~Wulz
% ...
\vskip\cmsinstskip
% ...
1:~~Also at CERN, European Organization for Nuclear Research, Geneva, Switzerland\\
% ...


%% file: EXO-11-013_temp.bbl
\providecommand{\href}[2]{#2}\begingroup\raggedright\begin{thebibliography}{10}%
\makeatletter
\providecommand{\hrefCMSnoop }[0]{\@secondoftwo}%
\makeatother

\bibitem{Pamela-positron}
\hrefCMSnoop {} {O.~Adriani {et~al.}, ``Observation of an anomalous positron
  abundance in the cosmic radiation'',} \textit{ Nature} \textbf{ 458} (2009)
  607. \href{http://dx.doi.org/10.1038/nature07942}{\texttt{
  doi:10.1038/nature07942}}.

\bibitem{Arkani-Hamed}
\hrefCMSnoop {} {N.~Arkani-Hamed {et~al.}, ``A Theory of Dark Matter'',}
  \textit{ Phys. Rev.} \textbf{ D79} (2009) 015014.
  \href{http://dx.doi.org/10.1103/PhysRevD.79.015014}{\texttt{
  doi:10.1103/PhysRevD.79.015014}}.

\bibitem{Dama}
\hrefCMSnoop {} {R.~Bernabei {et~al.}, ``New results from DAMA/LIBRA'',}
  \textit{ Eur. Phys. J.} \textbf{ C67} (2010) 39.
  \href{http://dx.doi.org/10.1140/epjc/s10052-010-1303-9}{\texttt{
  doi:10.1140/epjc/s10052-010-1303-9}}.

\bibitem{Finkbeiner}
\hrefCMSnoop {} {D.~P. Finkbeiner and N.~Weiner, ``Exciting dark matter and the
  INTEGRAL/SPI 511 keV signal'',} \textit{ Phys. Rev.} \textbf{ D76} (2007)
  083519. \href{http://dx.doi.org/10.1103/PhysRevD.76.083519}{\texttt{
  doi:10.1103/PhysRevD.76.083519}}.

\bibitem{Alves}
\hrefCMSnoop {} {D.~S.~M. Alves {et~al.}, ``Composite inelastic dark matter'',}
  \textit{ Phys. Lett.} \textbf{ B692} (2010) 323.
  \href{http://dx.doi.org/10.1016/j.physletb.2010.08.006}{\texttt{
  doi:10.1016/j.physletb.2010.08.006}}.

\bibitem{mediator-lepton-coupling2}
\hrefCMSnoop {} {P.~J. Fox and E.~Poppitz, ``Leptophilic Dark Matter'',}
  \textit{ Phys. Rev.} \textbf{ D79} (2009) 083528.
  \href{http://dx.doi.org/10.1103/PhysRevD.79.083528}{\texttt{
  doi:10.1103/PhysRevD.79.083528}}.

\bibitem{BaiHan}
\hrefCMSnoop {} {Y.~Bai and Z.~Han, ``Measuring the Dark Force at the LHC'',}
  \textit{ Phys. Rev. Lett.} \textbf{ 103} (2009) 051801.
  \href{http://dx.doi.org/10.1103/PhysRevLett.103.051801}{\texttt{
  doi:10.1103/PhysRevLett.103.051801}}.

\bibitem{Ruderman}
\hrefCMSnoop {} {J.~T. Ruderman {et~al.}, ``Non-abelian dark sectors and their
  collider signatures'',} \textit{ JHEP} \textbf{ 04} (2009) 014.
  \href{http://dx.doi.org/10.1088/1126-6708/2009/04/014}{\texttt{
  doi:10.1088/1126-6708/2009/04/014}}.

\bibitem{Strassler}
\hrefCMSnoop {} {M.~Strassler and K.~Zurek, ``Echoes of a hidden valley at
  hadron colliders'',} \textit{ Phys. Lett.} \textbf{ B651} (2007) 374.
  \href{http://dx.doi.org/10.1016/j.physletb.2007.06.055}{\texttt{
  doi:10.1016/j.physletb.2007.06.055}}.

\bibitem{D01}
\hrefCMSnoop {} {{D\O{} Collaboration}, ``Search for Dark Photons from
  Supersymmetric Hidden Valleys'',} \textit{ Phys. Rev. Lett.} \textbf{ 103}
  (2009) 081802.
  \href{http://dx.doi.org/10.1103/PhysRevLett.103.081802}{\texttt{
  doi:10.1103/PhysRevLett.103.081802}}.

\bibitem{D02}
\hrefCMSnoop {} {{D\O{} Collaboration}, ``Search for Events with Leptonic Jets
  and Missing Transverse Energy in $p\bar{p}$ collisions at $\sqrt{s}=1.96$
  TeV'',} \textit{ Phys. Rev. Lett.} \textbf{ 105} (2010) 211802.
  \href{http://dx.doi.org/10.1103/PhysRevLett.105.211802}{\texttt{
  doi:10.1103/PhysRevLett.105.211802}}.

\bibitem{Belle}
\hrefCMSnoop {} {{Belle Collaboration}, ``Search for a Low Mass Particle
  Decaying into $\mu^{+}\mu^{-}$ in $B^{0}\to K^{*0}X$ and $B^{0}\to\rho^{0}X$
  at Belle'',} \textit{ Phys. Rev. Lett.} \textbf{ 105} (2010) 091801.
  \href{http://dx.doi.org/10.1103/PhysRevLett.105.091801}{\texttt{
  doi:10.1103/PhysRevLett.105.091801}}.

\bibitem{babar-low-ma}
\hrefCMSnoop {} {{BABAR Collaboration}, ``Search for Dimuon Decays of a Light
  Scalar Boson in Radiative Transitions Upsilon $\to \gamma A^{0}$'',} \textit{
  Phys. Rev. Lett.} \textbf{ 103} (2009) 081803.
  \href{http://dx.doi.org/10.1103/PhysRevLett.103.081803}{\texttt{
  doi:10.1103/PhysRevLett.103.081803}}.

\bibitem{Aleph}
\hrefCMSnoop {} {{ALEPH Collaboration}, ``Search for neutral Higgs bosons
  decaying into four taus at LEP2'',} \textit{ JHEP} \textbf{ 05} (2010) 049.
  \href{http://dx.doi.org/10.1007/JHEP05(2010)049}{\texttt{
  doi:10.1007/JHEP05(2010)049}}.

\bibitem{CMS}
\hrefCMSnoop {} {{CMS Collaboration}, ``The CMS experiment at the CERN LHC'',}
  \textit{ JINST} \textbf{ 03} (2008) S08004.
  \href{http://dx.doi.org/10.1088/1748-0221/3/08/S08004}{\texttt{
  doi:10.1088/1748-0221/3/08/S08004}}.

\bibitem{Bernstein}
\hrefCMSnoop {} {S.~Bernstein, ``D\'{e}monstration du th\'{e}or\`{e}me de
  Weierstrass fond\'{e}e sur le calcul des probabilities'',} \textit{ Comm.
  Soc. Math. Kharkov} \textbf{ 13} (1912) 1.

\bibitem{crystal-ball}
\href {http://www.slac.stanford.edu/pubs/slacreports/slac-r-236.html} {M.~J.
  Oreglia, ``A study of the reactions $\psi^\prime \to \gamma \gamma \psi$''}.
\newblock PhD thesis, Stanford University, 1980.
\newblock {SLAC} Report {SLAC-R-236}, see Appendix {D}.

\bibitem{cteq6}
\hrefCMSnoop {} {P.~Nadolsky {et~al.}, ``Implications of CTEQ global analysis
  for collider observables'',} \textit{ Phys. Rev.} \textbf{ D78} (2008)
  013004. \href{http://dx.doi.org/10.1103/PhysRevD.78.013004}{\texttt{
  doi:10.1103/PhysRevD.78.013004}}.

\bibitem{nnpdf}
\hrefCMSnoop {} {R.~D. Ball {et~al.}, ``A first unbiased global NLO
  determination of parton distributions and their uncertainties'',} \textit{
  Nucl. Phys.} \textbf{ B838} (2010) 136.
  \href{http://dx.doi.org/10.1016/j.nuclphysb.2010.05.008}{\texttt{
  doi:10.1016/j.nuclphysb.2010.05.008}}.

\bibitem{mstw}
\hrefCMSnoop {} {A.~D. Martin {et~al.}, ``Parton distributions for the LHC'',}
  \textit{ Eur. Phys. J} \textbf{ C63} (2009) 189.
  \href{http://dx.doi.org/10.1140/epjc/s10052-009-1072-5}{\texttt{
  doi:10.1140/epjc/s10052-009-1072-5}}.

\end{thebibliography}\endgroup
